\documentclass[aps,prl,showpacs,twocolumn,superscriptaddress]{revtex4-1}

\usepackage[pdftex]{graphicx} 
\usepackage{natbib}
\usepackage{booktabs}
\usepackage{color}
\usepackage{url}
\usepackage{float}
\usepackage{amsmath}

\newcommand{\SCOC}{Sr$_2$CuO$_2$Cl$_2$}

\newcommand{\one}{Fig.~\ref{f1}}
\newcommand{\two}{Fig.~\ref{f2}}
\newcommand{\three}{Fig.~\ref{f3}}

\begin{document}
	
\title{Anomalous mirror symmetry breaking in a model insulating cuprate {\SCOC}}

\author{A. de la Torre}
\affiliation{Department of Physics, California Institute of Technology, Pasadena, CA 91125, USA}
\affiliation{Institute for Quantum Information and Matter, California Institute of Technology, Pasadena, CA 91125, USA}
\author{K. L. Seyler}
\affiliation{Department of Physics, California Institute of Technology, Pasadena, CA 91125, USA}
\affiliation{Institute for Quantum Information and Matter, California Institute of Technology, Pasadena, CA 91125, USA}
\author{L. Zhao}
\affiliation{Department of Physics, University of Michigan, 450 Church Street, Ann Arbor, Michigan 48109, USA}
\author{S. Di Matteo}
\affiliation{Universite de Rennes, CNRS UMR 6251 “Institut de Physique de Rennes”, F-35708 Rennes, France}
\author{M. S. Scheurer}
\affiliation{Department of Physics, Harvard University, Cambridge, MA 02138, USA}
\author{Y. Li}
\affiliation{School of Physics and Astronomy, University of Minnesota, Minneapolis, Minnesota 55455, USA}
\author{B. Yu}
\affiliation{School of Physics and Astronomy, University of Minnesota, Minneapolis, Minnesota 55455, USA}
\author{M. Greven}
\affiliation{School of Physics and Astronomy, University of Minnesota, Minneapolis, Minnesota 55455, USA}
\author{S. Sachdev}
\affiliation{Department of Physics, Harvard University, Cambridge, MA 02138, USA}
\author{M. R. Norman}
\affiliation{Materials Science Division, Argonne National Laboratory, Argonne, IL 60439, USA}
\author{D. Hsieh}
\affiliation{Department of Physics, California Institute of Technology, Pasadena, CA 91125, USA}
\affiliation{Institute for Quantum Information and Matter, California Institute of Technology, Pasadena, CA 91125, USA}

\date{\today}

\maketitle

\textbf{Understanding the complex phase diagram of cuprate superconductors is an outstanding challenge \cite{Keimer2015a}. The most actively studied questions surround the nature of the pseudogap and strange metal states and their relationship to superconductivity \cite{Norman2005, Fradkin2015, Proust2019}. In contrast, there is general agreement that the low energy physics of the Mott insulating parent state is well captured by a two-dimensional spin $S$ = 1/2 antiferromagnetic (AFM) Heisenberg model \cite{Lee2006}. However, recent observations of a large thermal Hall conductivity in several parent cuprates appear to defy this simple model and suggest proximity to a magneto-chiral state that breaks all mirror planes perpendicular to the CuO$_2$ layers \cite{Grissonnanche2019,Grissonnanche2020,Boulanger2020,Han2019,Samajdar2019}. Here we use optical second harmonic generation to directly resolve the point group symmetries of the model parent cuprate {\SCOC}. We report evidence of an order parameter $\Phi$ that breaks all perpendicular mirror planes and is consistent with a magneto-chiral state in zero magnetic field. Although $\Phi$ is clearly coupled to the AFM order parameter, we are unable to realize its time-reversed partner ($-\Phi$) by thermal cycling through the AFM transition temperature ($T_{\textrm{N}}$ $\approx$ 260 K) or by sampling different spatial locations. This suggests that $\Phi$ onsets above $T_{\textrm{N}}$ and may be relevant to the mechanism of pseudogap formation.}

The single layer oxychloride cuprate {\SCOC} is an ideal parent material for studying subtle magnetic symmetry breaking effects because $T_{\textrm{N}}$ is easily accessible and because its crystallographic structure (space group 139, $I4/mmm$) has exceptionally high symmetry. As shown in {\one}(a), {\SCOC} consists of tetragonal CuO$_2$ planes separated by Sr$_2$Cl$_2$ buffer layers. Unlike other commonly studied cuprates \cite{Proust2019}, no long-range tilting/rotation of the CuO$_4$Cl$_2$ octahedra or structural modulations are present either above or below $T_{\textrm{N}}$ in {\SCOC} \cite{Grande1975,Miller1990}. This structure results in a very small magnetic anisotropy \cite{Cuccoli2003,Suh1996,Katsumata2001,Greven1995}, extremely weak interlayer coupling due to frustration of the interlayer exchange \cite{Greven1995}, and an absence of Dzyaloshinskii-Moriya (DM) interactions, making {\SCOC} a model 2D Heisenberg antiferromagnet.

\begin{figure}[!ht]
	\includegraphics[width=8.5cm]{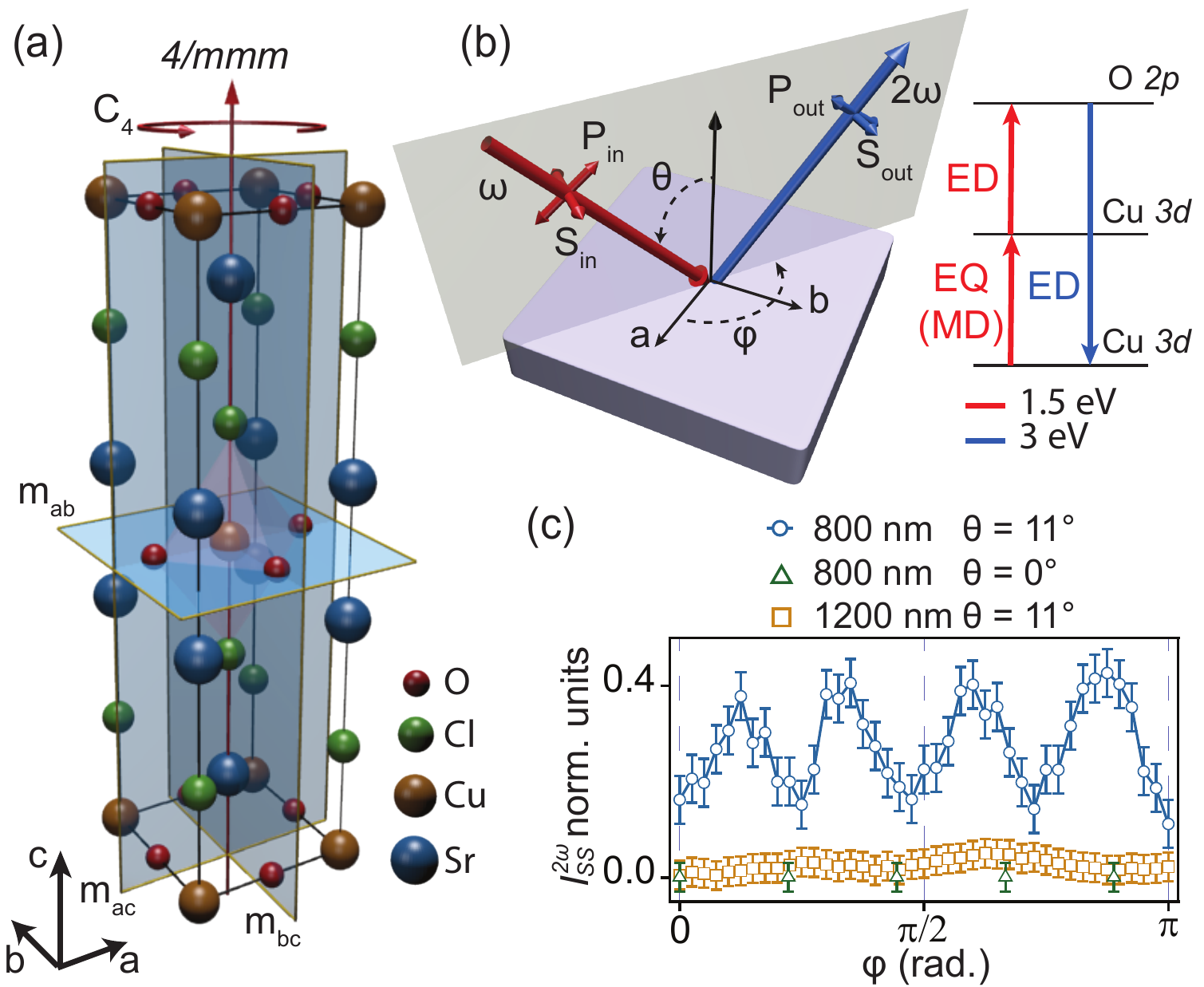} %
	\caption{(a) Unit cell of {\SCOC} showing the generators (m$_{bc}$, m$_{ac}$, m$_{ab}$ and C$_4$) of the tetragonal 4/\textit{mmm} point group. (b) Schematic of the RA-SHG setup. The scattering plane angle ($\varphi$), incidence angle ($\theta$) and input and output electric field polarization (S or P) are varied in our experiment. The states involved in the SHG processes discussed in the main text are shown in the inset. (c) $\varphi$-dependence (0 $\rightarrow$ $\pi$ shown for clarity) of the SHG intensity from {\SCOC} at different incident angles and wavelengths measured at $T = 300$ K in the S$_{\textrm{in}}$S$_{\textrm{out}}$ (abbreviated SS) channel. Fluences and counting times were held constant. Error bars represent the statistical uncertainty. The 800 nm ($\sim$1.5 eV) $\theta = 11^{\circ}$ curve is offset vertically for clarity.}
	\label{f1}
\end{figure}

\begin{figure*}[!ht]
	\includegraphics[width=17.5cm]{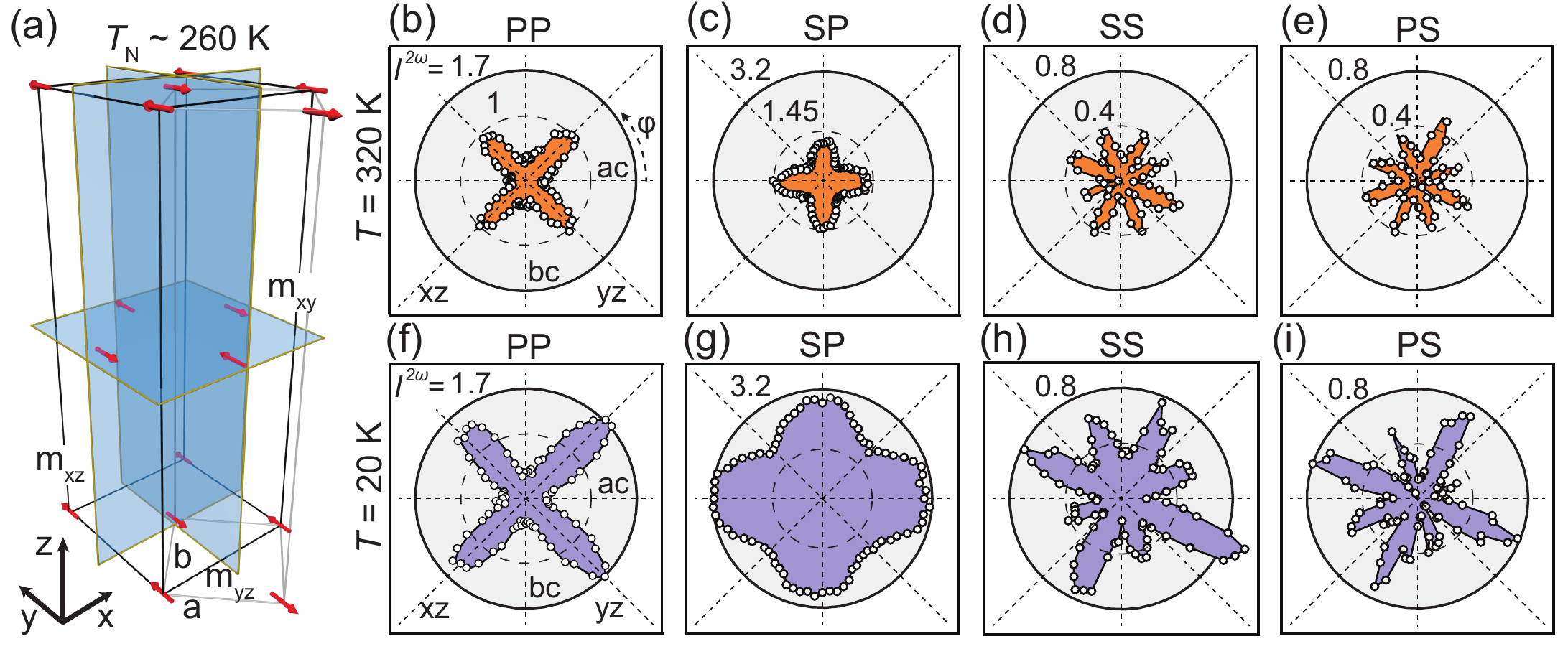} %
	\caption{(a) AFM structure of {\SCOC} showing the spatial symmetry generators (m$_{yz}$, m$_{xz}$ and m$_{xy}$) of its orthorhombic $mmm1'$ point group. Moments are aligned along the $y$-axis $[\bar{1}10]$ direction. Although time-reversal symmetry is locally broken at each Cu site, it is restored upon translation by an in-plane lattice vector. Therefore the time-reversal operator $1'$ is a symmetry of the point group. RA data for four polarization channels measured at (b)-(e) $T = 320$ K (above $T_{\textrm{N}}$) and (f)-(i) $T = 20$ K. All data sets are normalized to the maximum intensity in the PP channel at $T$ = 320 K. There is larger statistical noise in the S-output channels due to their extremely weak intensity.}
	\label{f2}
\end{figure*}

Second harmonic generation (SHG) is governed by high rank ($>$ 2) nonlinear optical susceptibility tensors. Therefore it is much more sensitive to the crystallographic and magnetic point group symmetries of a crystal compared to typical linear optical responses \cite{Fiebig2005}. To resolve the SHG tensor structures of {\SCOC}, we performed SHG rotational anisotropy (RA) and spectroscopy measurements in reflection mode from (001) cleaved single crystals \cite{Harter2015}. Figure~\ref{f1}(b) provides a depiction of our setup, showing all of the degrees of freedom used in our experiment.

We first examine the SHG response of {\SCOC} above $T_{\textrm{N}}$. The 4/$mmm$ point group respects inversion symmetry and so no bulk electric-dipole (ED) SHG is expected. However, surface ED SHG and higher multipole bulk processes such as electric quadrupole (EQ) SHG are allowed. These contributions are readily distinguished in the SS polarization channel [{\one}(b)] because ED SHG from the (001) surface is forbidden by symmetry whereas the bulk EQ SHG intensity should scale as $I^{2\omega}_{\textrm{SS}}\propto\sin^2\theta \sin^24\varphi$ \cite{SM}, where $\theta$ is the angle of incidence and $\varphi$ is the scattering plane angle. As shown in {\one}(c), the SHG intensity from {\SCOC} (001) measured using 800 nm incident light vanishes at normal incidence ($\theta = 0^{\circ}$) but becomes finite, albeit very weak ($\sim10^4$ times weaker than GaAs \cite{SM}), at oblique incidence ($\theta = 11^{\circ}$) with a clear $\sin^24\varphi$ dependence. This indicates predominant sensitivity to bulk EQ SHG and consistency with a 4/$mmm$ point group. Moreover, we find no detectable SHG using 1200 nm (1 eV) incident light [{\one}(c)]. Given that previous optical conductivity results on {\SCOC} show an ED forbidden Cu $d$-$d$ excitation feature near 1.5 eV (800 nm) and a broad band of ED allowed O $2p$ to Cu $3d$ charge transfer excitations around 2-3 eV \cite{Choi1999}, we attribute our observed SHG signal to a two-photon process resonantly enhanced by an EQ $d$-$d$ transition [inset {\one}(b)].

A comparison of RA patterns obtained above and below $T_{\textrm{N}}$ under all different polarization geometries is shown in {\two}. The $T$ = 320 K data exhibit four-fold (C$_4$) rotational symmetry about the $c$-axis and mirror symmetry about the $ac$ and $bc$ planes as well as the diagonal $xz$ and $yz$ planes, which are all elements of the 4/$mmm$ point group. These data are again consistent with bulk EQ SHG and incompatible with surface ED SHG because the latter is forbidden in the SS and PS channels and is $\varphi$-independent in the PP and SP channels \cite{SM}. At $T$ = 20 K, the SHG intensity is generally larger than at $T$ = 320 K in all polarization channels. However, while the PP and SP patterns retain all the symmetries of the $T$ = 320 K data, the PS and SS patterns retain C$_4$ but break the $ac$, $bc$, $xz$ and $yz$ mirror planes, manifested via a change from uniform to alternating lobe intensity.

\begin{figure}[htb]
	\includegraphics[width=8cm]{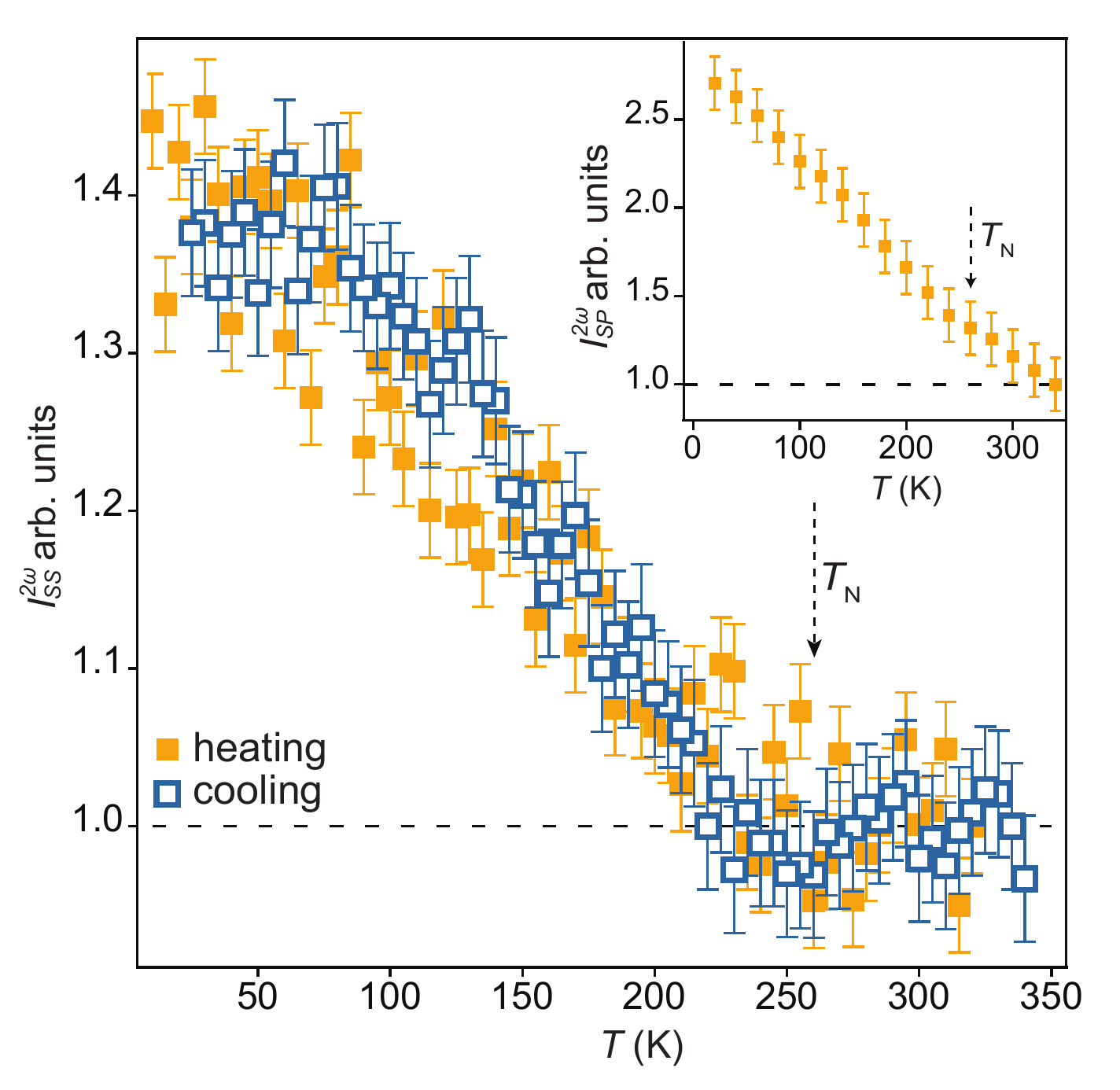} %
	\caption{SHG intensity measured in the SS channel at $\varphi=3\pi/8$ upon heating (orange symbols) and cooling (blue symbols). Data are normalized to the $T = 320$ K value. Inset shows analogous data in the SP channel at $\varphi=\pi/8$. Error bars represent the standard deviation over four independent measurements. See \cite{SM} for PP and PS data.}
	\label{f3}
\end{figure}

As shown in Figure 3, the change in the RA patterns observed in the S-output channels onsets near $T_{\textrm{N}}$ and gradually becomes more pronounced upon cooling. The absence of thermal hysteresis is consistent with a continuous phase transition at $T_{\textrm{N}}$. Although this suggests that the change in RA patterns is correlated with AFM ordering, it cannot explain the observed symmetry lowering. Below $T_{\textrm{N}}$, spins on the Cu sites of {\SCOC} adopt a collinear AFM arrangement characterized by an ordering wave vector ($\frac{1}{2}$,$\frac{1}{2}$,0), with moments aligned along the $[\bar{1}10]$ axis \cite{Vaknin1990}. This spin structure is described by an orthorhombic magnetic point group $mmm1'$, whose generators consist of three mirror operators about the $xy$, $xz$, and $yz$ planes as well as the time-reversal operator $1'$ [{\two}(a)]. The absence of $xz$ and $yz$ mirror symmetries in the RA patterns [{\two}(h) \& (i)] therefore shows that the SHG response does not directly couple to the AFM order parameter ($\Psi$). This is expected because neighboring Cu sites are structurally equivalent and related purely by time-reversal below $T_{\textrm{N}}$, guaranteeing cancellation of any SHG process that is proportional to the local moment. It is possible for the SHG response to couple to $\Psi^2$ via magnetostriction \cite{Fiebig2005}, but the associated crystallographic point group is orthorhombic (point group $mmm$) and preserves the $xz$ and $yz$ mirror symmetries \cite{SM}. Averaging over 90$^{\circ}$ rotated orthorhombic domains would not lift these symmetries in the RA-SHG patterns.

\begin{figure*}[htb]
	\includegraphics[width=13cm]{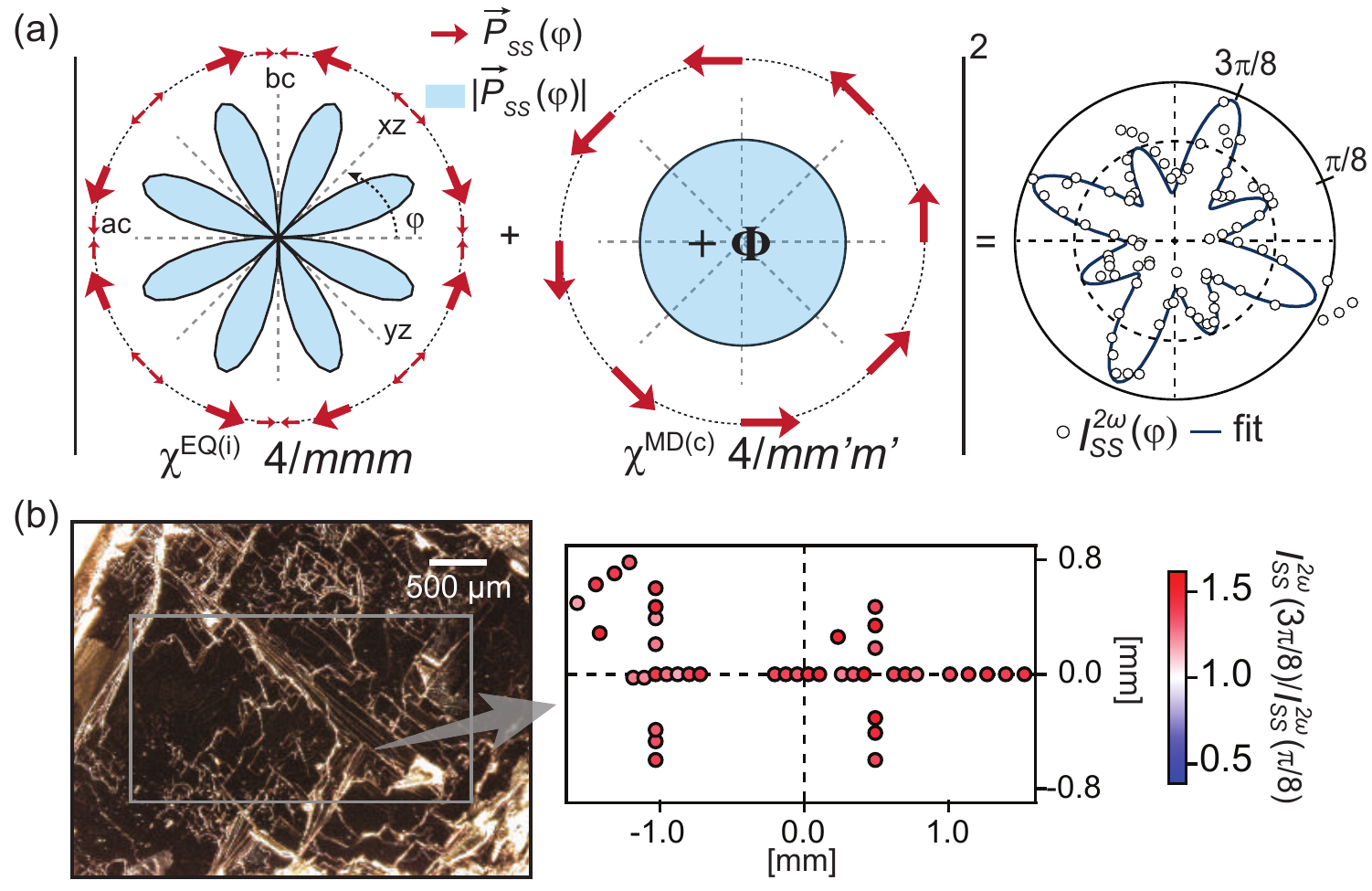} %
	\caption{(a) Illustration of how the low temperature RA pattern in the SS channel can be produced by a coherent superposition of $\chi^{\textrm{EQ}(i)}$ (4/$mmm$) and $\chi^{\textrm{MD}(c)}$ (4/$mm'm'$) processes. The red arrows denote the S-input light induced electric polarization projected along the S-output direction ($\vec{P}_{SS}$) at each $\varphi$. The polar plots show the $\varphi$ dependence of $|\vec{P}_{SS}|$. A fit of this model (blue lines) to the RA data (circles) in the SS channel measured at $T$ = 20 K is shown on the right. The best fit was obtained using an optical phase difference of 102$^{\circ}$ between the EQ and MD processes. (b) Optical micrograph of the cleaved (001) surface of a {\SCOC} single crystal. RA-SHG patterns were measured in the SS channel at $T$ = 80 K at the locations marked by colored circles within the gray box. The color scale encodes the intensity ratio between two adjacent lobes at $\varphi = 3\pi/8$ and $\pi/8$ extracted from fits [see panel (a)]. The slight spatial-dependence of this ratio is primarily due to systematic noise.}
	\label{f4}
\end{figure*}

We first consider possible structural mechanisms for the observed symmetry breaking. The removal of vertical mirror planes lowers the crystallographic point group of {\SCOC} from 4/$mmm$ to 4/$m$. This could arise, for example, through a staggered rotation and distortion of the CuO$_4$Cl$_2$ octahedra analogous to Sr$_2$IrO$_4$ \cite{Ye2015}. However, x-ray and neutron diffraction studies find no evidence of such distortions either above or below $T_{\textrm{N}}$ \cite{Grande1975,Vaknin1990,Miller1990}. One can postulate that the 4/$m$ distortion only occurs at the surface. However, surface-sensitive He-atom scattering \cite{Farzaneh2005} and angle-resolved photoemission spectroscopy measurements \cite{Durr2000} on {\SCOC} show no signs of (001) surface reconstruction or unit cell enlargement. Even if the mirror planes are somehow broken at the surface, it still cannot explain our data because ED SHG from the surface of a 4/$m$ structure is forbidden in the SS channel \cite{SM}. Moreover, it is unclear physically how an orthorhombic AFM order parameter with $mmm1'$ symmetry can induce a secondary structural order parameter with tetragonal 4/$m$ symmetry.

An alternative explanation is that the vertical mirror planes are broken by a hitherto undetected magnetic order parameter ($\Phi$) that is coupled non-linearly to $\Psi$ and is described by a 4/$mm'm'$ magnetic point group. Here $m'$ denotes a combined mirror and time-reversal operation. Lower symmetry subgroups of 4/$mm'm'$ cannot be ruled out but are not necessary to explain our data. A magnetic state with point group 4/$mm'm'$ has $A_{2g}$ symmetry (inversion preserving) and transforms like the $z$-component of magnetization. Hence it is compatible with a ferroic ordering of odd-rank magnetic moments (dipole, octupole, etc \cite{SM}) along the $c$-axis and is consistent with a magneto-chiral state. Below, we survey the new SHG processes that can be induced by this magnetic state and interfere with the existing 4/$mmm$ crystallographic EQ SHG signal.

One possibility is that the new SHG response couples linearly to $\Phi$. This can occur through an EQ process of the form $P_i^{2\omega} = \chi^{\textrm{EQ}(c)}_{ijkl} E_j^{\omega}\nabla_{k}E_l^{\omega}$ or a magnetic-dipole (MD) process of the form $P_i^{2\omega} = \chi^{\textrm{MD}(c)}_{ijk} E_j^{\omega}H_k^{\omega}$, where $E$ and $H$ are the electric and magnetic fields of the incident light at frequency $\omega$, $P$ is the induced electric polarization at the second harmonic, the $\chi$'s are $c$-type (time-reversal odd) susceptibility tensors that are invariant under the symmetries of the 4/$mm'm'$ point group, and the $i,j,k$ indices run through the $x,y,z$ coordinates. Another possibility is that the new SHG response couples to $\Phi^2$ via magnetostriction, which would be governed by $i$-type (time-reversal even) EQ and MD susceptibility tensors respecting the 4/$m$ point group symmetries. We calculated expressions for the RA intensity in all of these cases \cite{SM}. Among the aforementioned processes, the $\chi^{\textrm{MD}(c)}$ process stands out because $I^{2\omega}_{\textrm{PP}} = I^{2\omega}_{\textrm{SP}} = 0$, $I^{2\omega}_{\textrm{PS}} \propto |\chi^{\textrm{MD}(c)}_{xzx} \sin\theta|^2$ and $I^{2\omega}_{\textrm{SS}} \propto |\chi^{\textrm{MD}(c)}_{xxz} \sin\theta|^2$. Attributing the SHG signal that onsets below $T_{\textrm{N}}$ predominantly to this MD process would explain why we only observe mirror symmetry breaking below $T_{\textrm{N}}$ in the S-output channels and not the P-output channels [Figs 2 \& 3]. It is also consistent with the absence of any normal incidence SHG signal below $T_{\textrm{N}}$ \cite{SM}.

The quasi-linear intensity increase in the P-output channels observed upon cooling [{\three} inset] can be attributed to non-uniform changes in the crystallographic $\chi^{\textrm{EQ}(i)}$ elements \cite{Ron2019}. This likely arises from the anisotropic thermal contraction of the {\SCOC} lattice. As shown by neutron diffraction, between $T$ = 300 K and 25 K the $c$-axis contracts by $\sim$0.53\% as compared to only $\sim$0.23\% for the $a$-axis \cite{Miller1990}. Therefore one expects a stronger temperature dependence for out-of-plane ($E \parallel c$) compared to in-plane ($E \perp c$) excitations. The fact that only the P-output channels are sensitive to out-of-plane excitations (i.e. $\chi^{\textrm{EQ}(i)}$ elements with $i$, $j$ or $l$ = $z$) can explain why they exhibit the more marked temperature-dependent background \cite{SM}.

Figure 4(a) illustrates how the observed symmetry breaking in the S-output channels can be generated by a new $\chi^{\textrm{MD}(c)}$ contribution below $T_{\textrm{N}}$. The SHG response above $T_{\textrm{N}}$ is governed by an $i$-type EQ susceptibility tensor that respects 4/$mmm$ symmetry and generates an RA pattern with eight lobes of equal intensity [{\two}(d) \& (e)]. Below $T_{\textrm{N}}$ a $c$-type MD susceptibility tensor that respects 4/$mm'm'$ symmetry emerges and generates a $\varphi$-independent contribution. Coherent superposition of these two processes yields oblique incidence RA intensities proportional to $|\alpha\sin4\varphi + \beta|^2$, where $\alpha = \chi^{\textrm{EQ}(i)}_{xyxy}-\chi^{\textrm{EQ}(i)}_{xxxx}+2\chi^{\textrm{EQ}(i)}_{xxyy}$ and $\beta$ = $\chi^{\textrm{MD}(c)}_{xzx}$ or $\chi^{\textrm{MD}(c)}_{xxz}$ for the PS and SS channels respectively. We note that in an absorbing medium, the tensor elements are generally complex. This allows interference between the EQ and MD contributions, resulting in an alternating lobe intensity and a lifting of nodes that is consistent with observations.

Finally we discuss the microscopic origins of $\Phi$. The simplest magnetic state with $A_{2g}$ symmetry is a $c$-axis oriented ferromagnet, which can arise from uniform canting of the Cu spins out of the $ab$ plane. But x-ray magnetic circular dichroism measurements on {\SCOC} detect no out-of-plane spin canting \cite{DeLuca2010}, in keeping with its reported structure that forbids DM interactions. Surface spin canting also seems unlikely given the absence of surface reconstruction \cite{Farzaneh2005,Durr2000}. Furthermore, in the spontaneous spin canting scenario, one might expect to find a spatial distribution of time-reversed domains characterized by $\pm\Phi$ that can also flip upon thermal cycling across $T_{\textrm{N}}$. Since $\chi^{\textrm{MD}(c)}_{xxz}$ couples linearly to $\Phi$, a change in the sign of $\Phi$ should invert the intensity ratio of adjacent lobes in the S-output RA patterns. We collected low temperature RA patterns in the SS channel at multiple locations spanning millimeters across a sample using an optical spot size of $\sim$40 $\mu$m. We deliberately selected a cleaved surface with many terrace steps to search for both lateral and $c$-axis domains. Figure~\ref{f4}(b) shows the intensity ratio of adjacent lobes $I^{2\omega}_{\textrm{SS}}(\varphi=3\pi/8)/I^{2\omega}_{\textrm{SS}}(\varphi=\pi/8)$ as a function of position. Surprisingly, this ratio stays greater than 1 at all measured locations and following multiple thermal cycles across $T$ = 320 K \cite{SM}. This indicates that $\Phi$ is locked to one sign, further arguing against a Cu spin canting scenario.

Instead, our results suggest that $\Phi$ may arise from an independent magnetic state that already exists above $T$ = 320 K and intrinsically forms large single domains, or has its orientation pinned by extrinsic effects such as correlated structural defects. The order parameter $\Phi$ is merely enhanced at $T_{\textrm{N}}$ due to coupling to $\Psi$ so as to become more clearly detectable by SHG. Indeed, since $\Phi$ and $\Psi$ break independent symmetries, they are not constrained to have the same critical temperature and a coupling term of the form $\Phi^2\Psi^2$ is allowed by symmetry in the free energy. More complex intra-unit-cell spin or spin-lattice coupled arrangements with higher multipole moments \cite{Fechner2016}, intra-unit-cell orbital loop current configurations \cite{Varma1997,He2014,Scheurer2018} possibly associated with topological order \cite{Sachdev2003}, or long wavelength orbital magnetization density waves \cite{Dai2019} that break vertical mirror planes are plausible candidates for $\Phi$. We note that the AFM form factor of {\SCOC} measured from neutron diffraction reportedly deviates from that of a free Cu$^{2+}$ ion \cite{Wang1990}. Resonant inelastic x-ray scattering measurements of its magnetic excitation spectrum also reveal the importance of further neighbor and four-spin ring exchange interactions \cite{Guarise2010,Plumb2014}. These data support the possibility of more exotic orders beyond the classical AFM state. Regardless of microscopic origin, the existence of $\Phi$ may be connected to several unexplained properties of the cuprate pseudogap state including a large chirality-induced thermal Hall conductivity \cite{Grissonnanche2019,Boulanger2020} (allowed by 4/$mm'm'$ but not $mmm1'$ \cite{Seeman2015}), the enhancement of vertical mirror symmetry breaking observed by THz and SHG polarimetry \cite{Lubashevsky2014,Zhao2017,SM}, and a polar Kerr rotation (allowed by 4/$mm'm'$) that can be magnetic field trained well above the pseudogap temperature \cite{Kapitulnik2008}.

\section{Acknowledgements}
We acknowledge helpful conversations with Dante Kennes, Steve Kivelson, Patrick Lee, Olexei Motrunich, Damjan Pelc and Kemp Plumb. We also thank Louis Taillefer and Ga\"{e}l Grissonnanche for sharing unpublished data. The SHG work is supported by an ARO PECASE award W911NF-17-1-0204. D.H. also acknowledges support for instrumentation from the David and Lucile Packard Foundation and from the Institute for Quantum Information and Matter (IQIM), an NSF Physics Frontiers Center (PHY-1733907). A.d.l.T. acknowledges support from the Swiss National Science Foundation through an Early Postdoc Mobility Fellowship (P2GEP2$\_$165044). K.L.S. acknowledges a Caltech Prize Postdoctoral Fellowship. S.S. acknowledges support from NSF grant DMR-2002850. M.S.S. acknowledges support from the German National Academy of Sciences Leopoldina through Grant LPDS 2016-12. M.R.N. was supported by the Materials Sciences and Engineering Division, Basic Energy Sciences, Office of Science, U.S. Department of Energy. The work at the University of Minnesota was funded by the U.S. Department of Energy through the University of Minnesota Center for Quantum Materials, under Grant No. DE-SC-0016371.

\section{Methods}

\subsection{Single crystal growth and preparation}
Single crystals of {\SCOC} were grown by standard methods described in Ref.\cite{Miller1990}. Crystals were pre-aligned using x-ray Laue diffraction, cleaved along their (001) surface in a N$_2$-rich environment to avoid air exposure, and then immediately pumped down to pressures below $10^{-7}$ torr.

\subsection{Second harmonic generation measurements}
The RA-SHG experiments were performed using a high-speed rotating scattering plane based setup \cite{Harter2015}. Incident laser light was delivered by a Ti:sapphire regenerative amplifier (pulse width 80 fs, repetition rate 100 kHz, center wavelength 800 nm) seeding an optical parametric amplifier. The light was focused to a spot size of $\sim$40 $\mu$m onto optically flat regions of the crystals. The infrared photon energies used in this study are below the charge gap of {\SCOC} and very weakly absorbed \cite{Perkins1993}. Therefore a minimum fluence of around 3 mJ/cm$^2$ was necessary to acquire RA-SHG patterns of reasonable quality using obliquely incident 800 nm light. All data presented were reproduced on multiple spots across multiple single crystals.



%
\newpage
\pagenumbering{gobble}
\begin{figure}
   \vspace*{-2cm}
   \hspace*{-2cm}
    \centering
    \includegraphics[page=1]{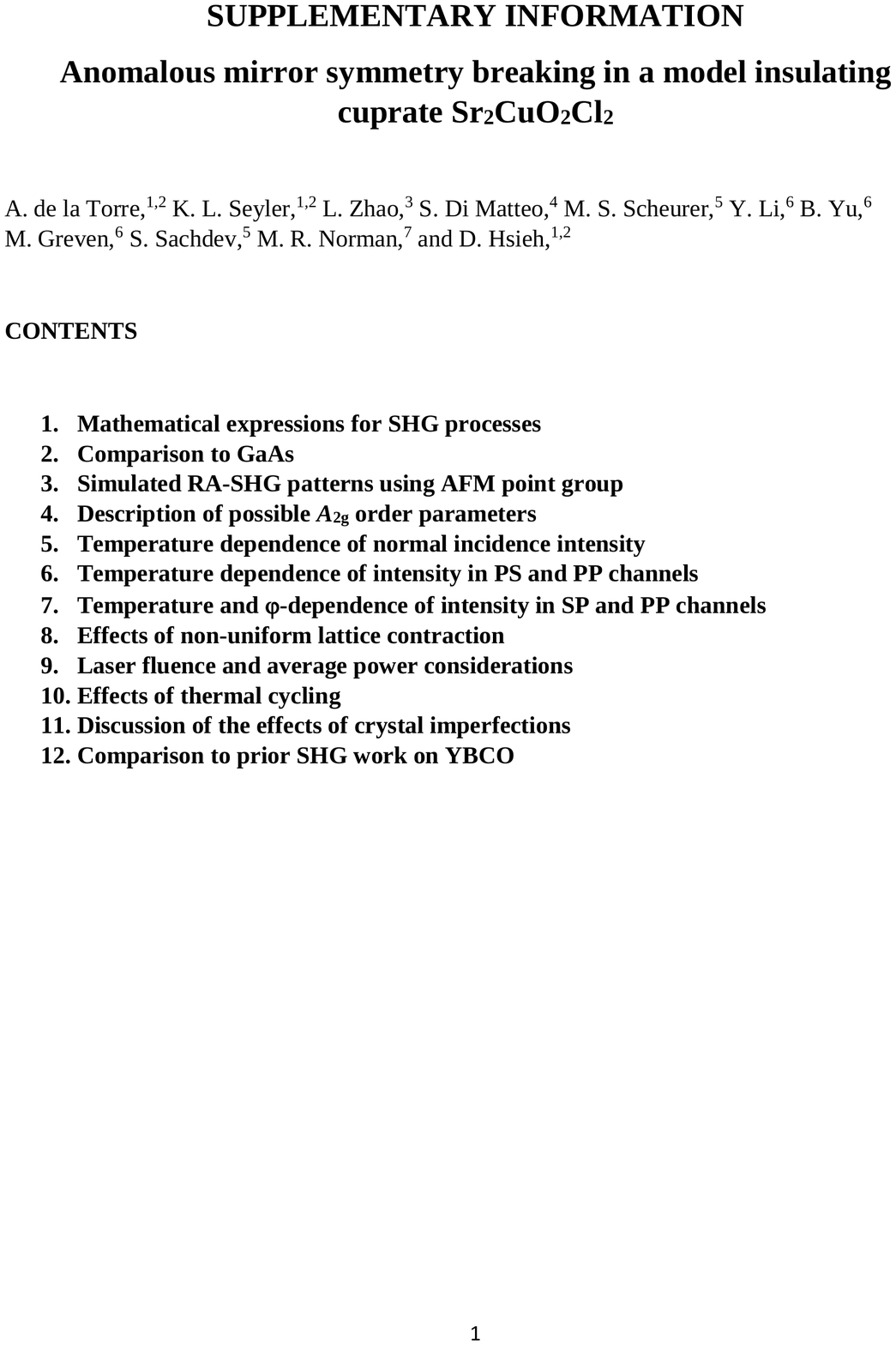}
    \caption{Caption}
    \label{fig:my_label}
\end{figure}
\begin{figure}
   \vspace*{-2cm}
   \hspace*{-2cm}
    \centering
    \includegraphics[page=2]{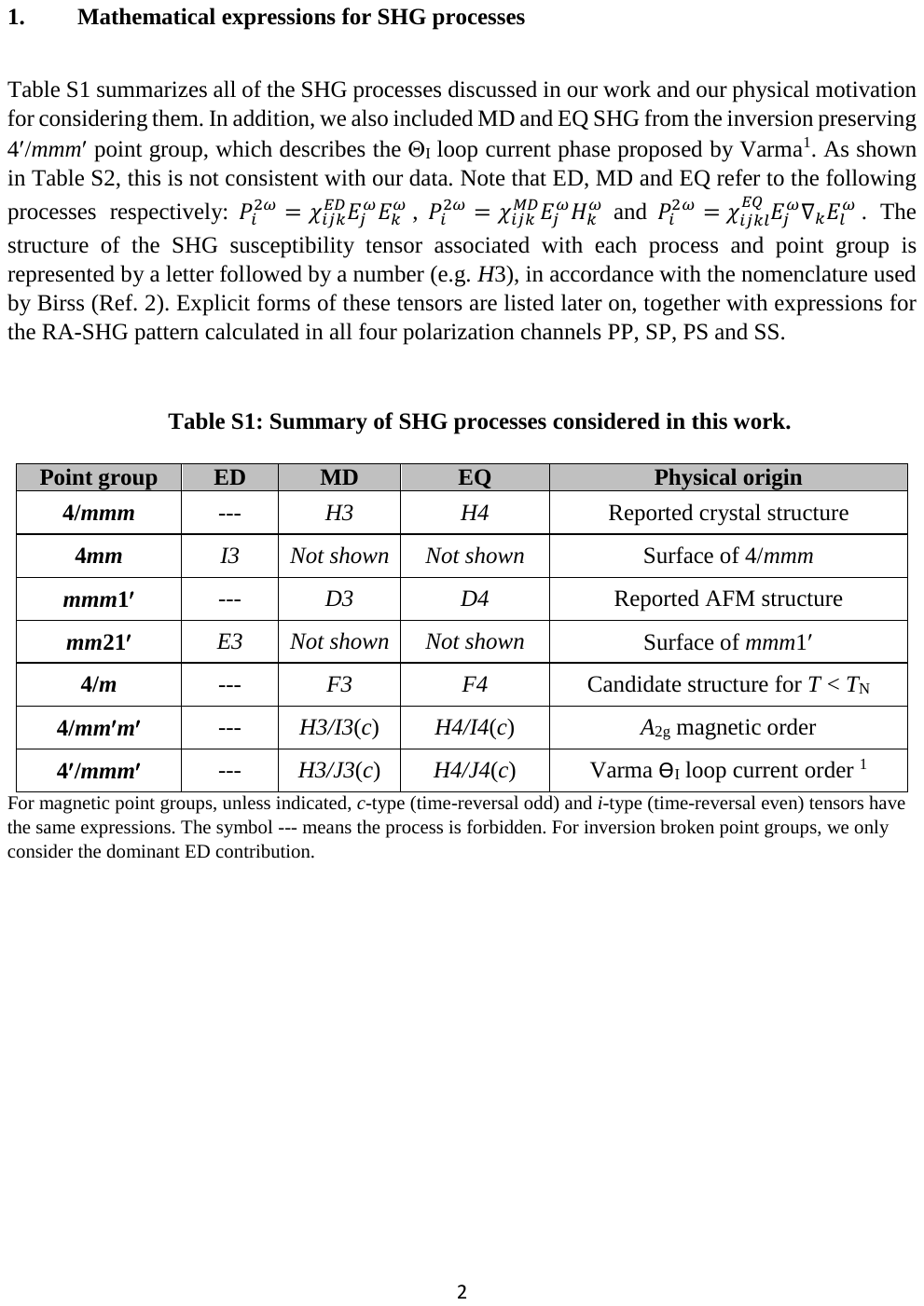}
    \caption{Caption}
    \label{fig:my_label}
\end{figure}
\begin{figure}
   \vspace*{-2cm}
   \hspace*{-2cm}
    \centering
    \includegraphics[page=3]{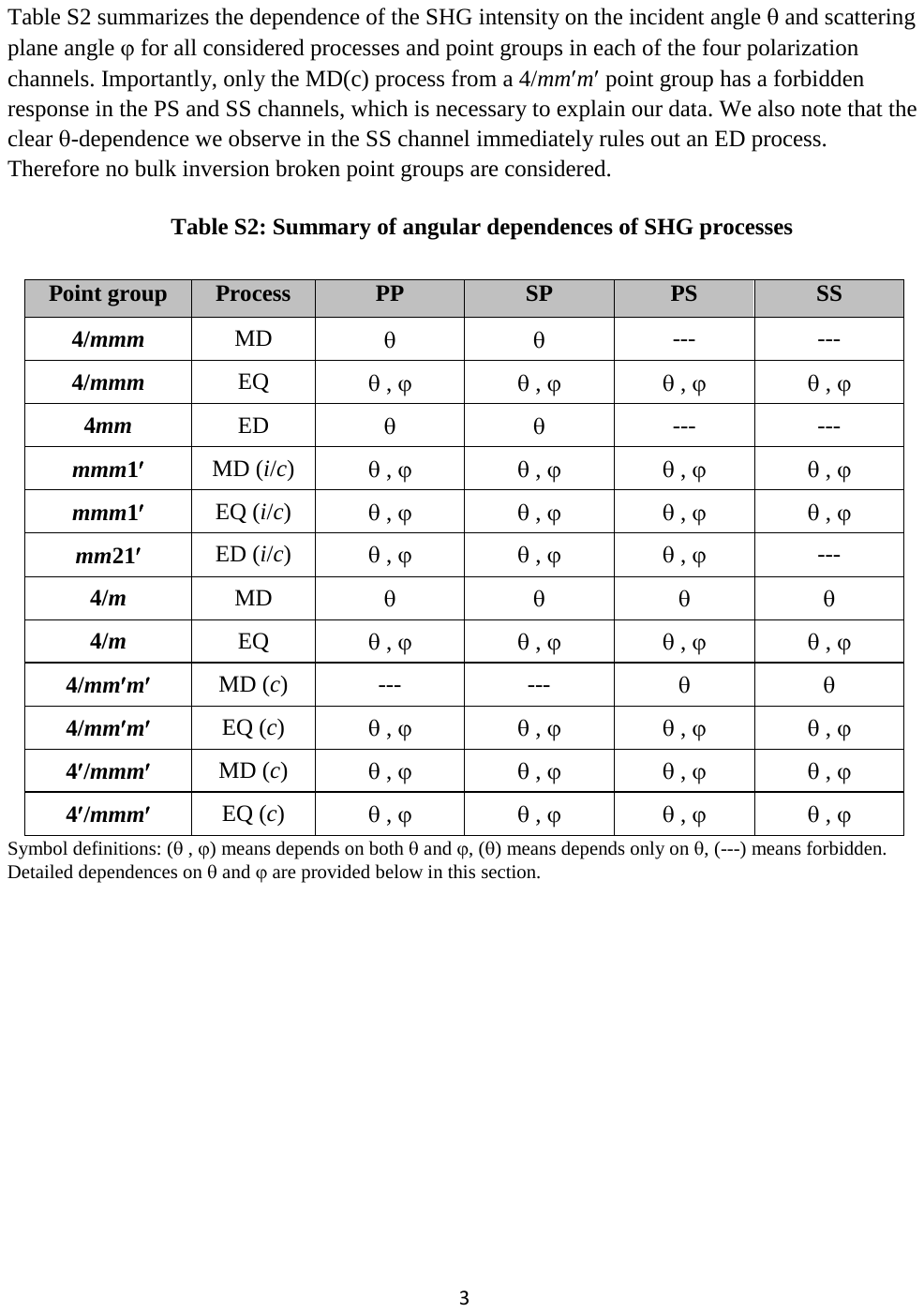}
    \caption{Caption}
    \label{fig:my_label}
\end{figure}
\begin{figure}
   \vspace*{-2cm}
   \hspace*{-2cm}
    \centering
    \includegraphics[page=4]{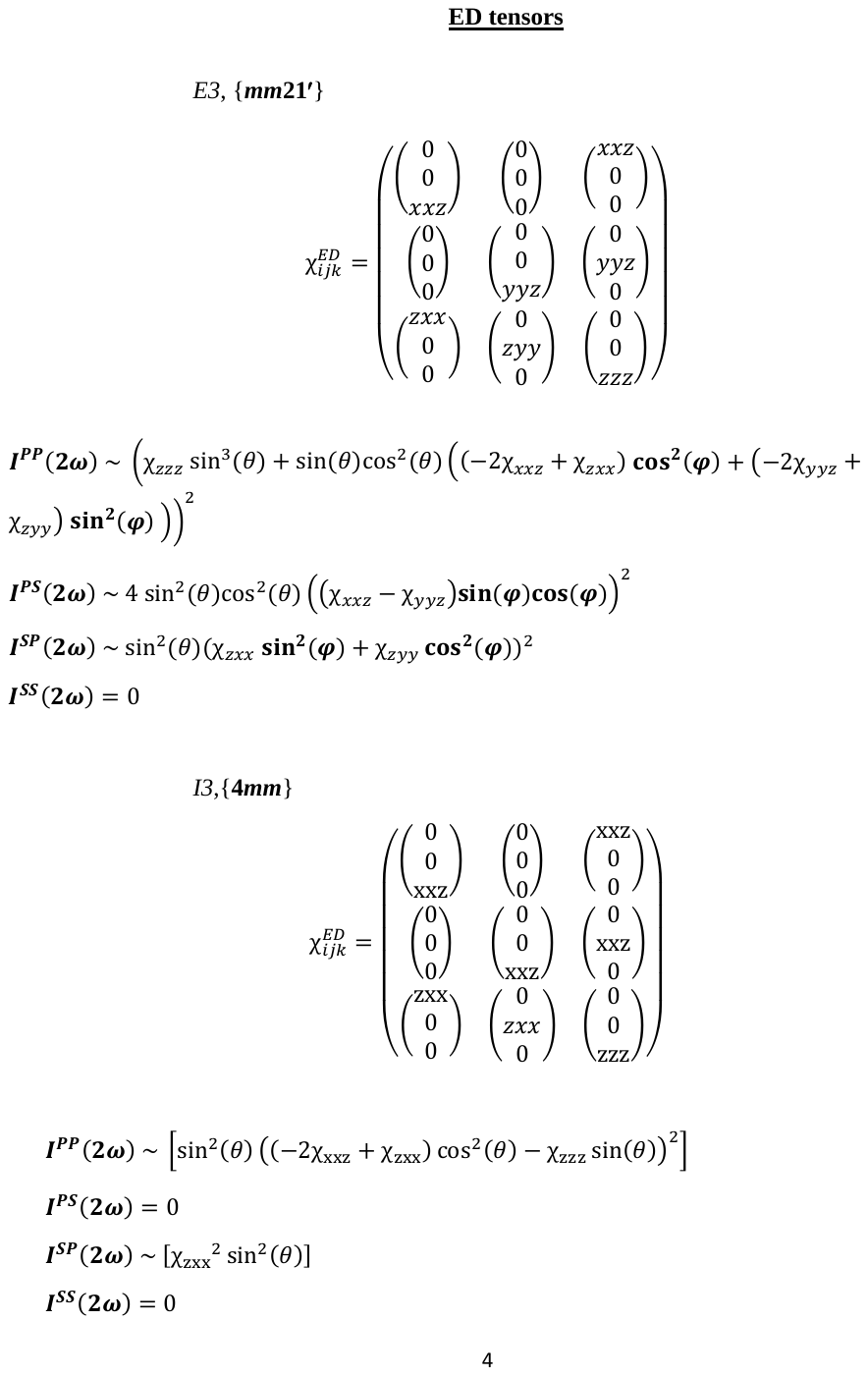}
    \caption{Caption}
    \label{fig:my_label}
\end{figure}
\begin{figure}
   \vspace*{-2cm}
   \hspace*{-2cm}
    \centering
    \includegraphics[page=5]{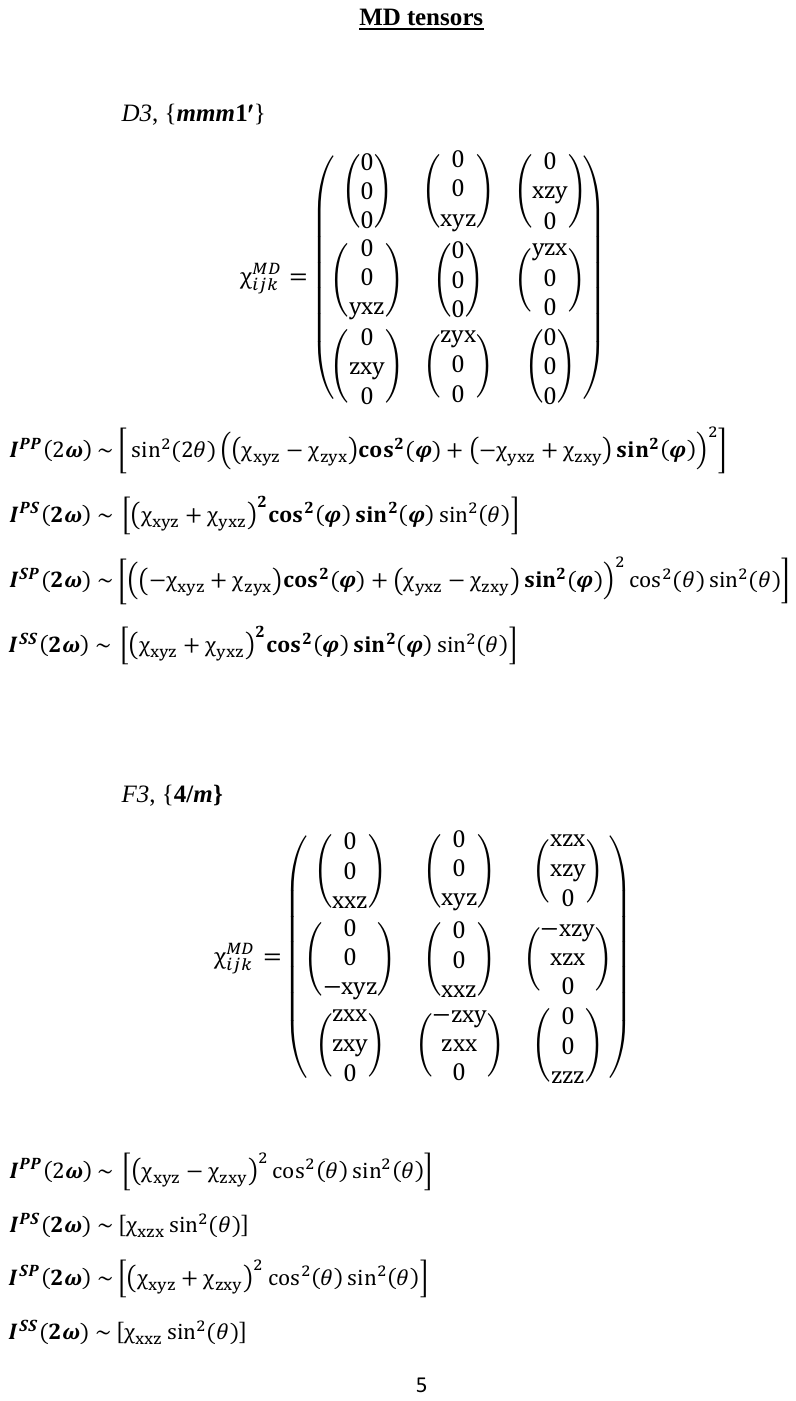}
    \caption{Caption}
    \label{fig:my_label}
\end{figure}
\begin{figure}
   \vspace*{-2cm}
   \hspace*{-2cm}
    \centering
    \includegraphics[page=6]{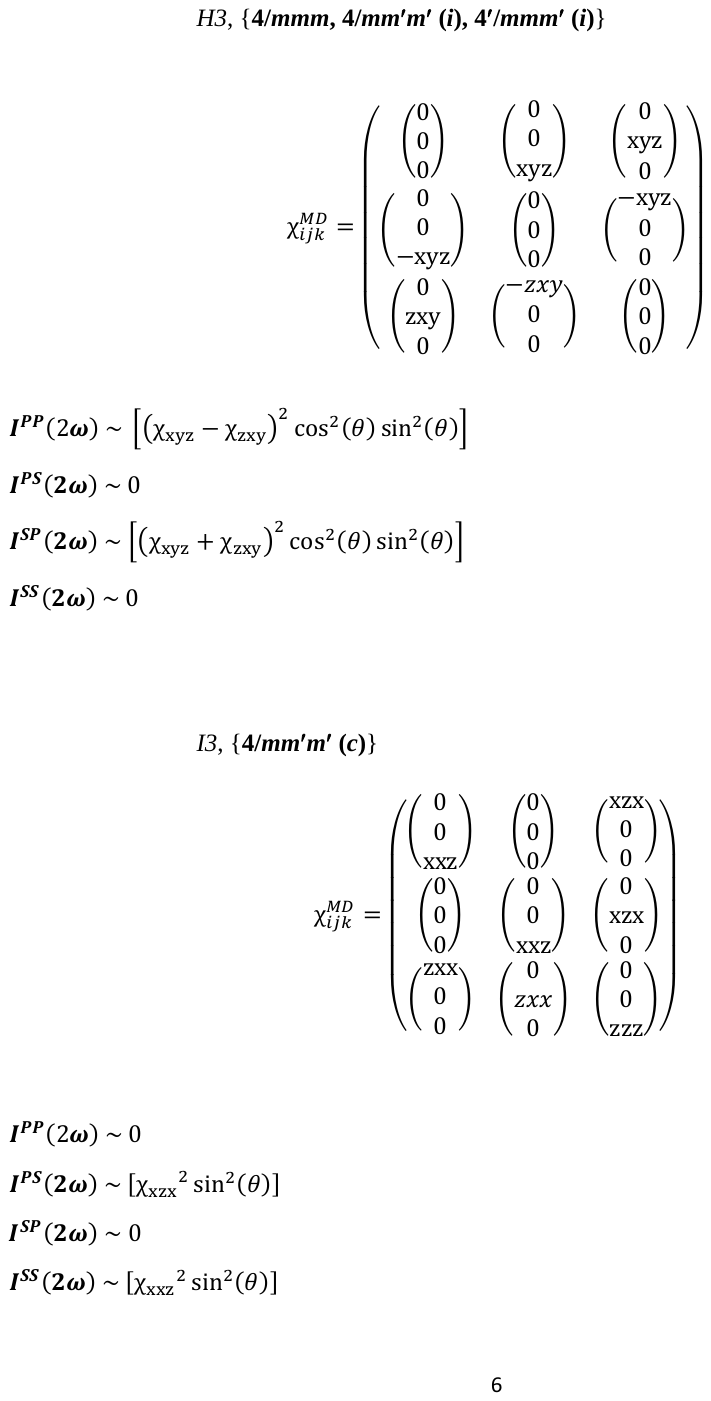}
    \caption{Caption}
    \label{fig:my_label}
\end{figure}
\begin{figure}
   \vspace*{-2cm}
   \hspace*{-2cm}
    \centering
    \includegraphics[page=7]{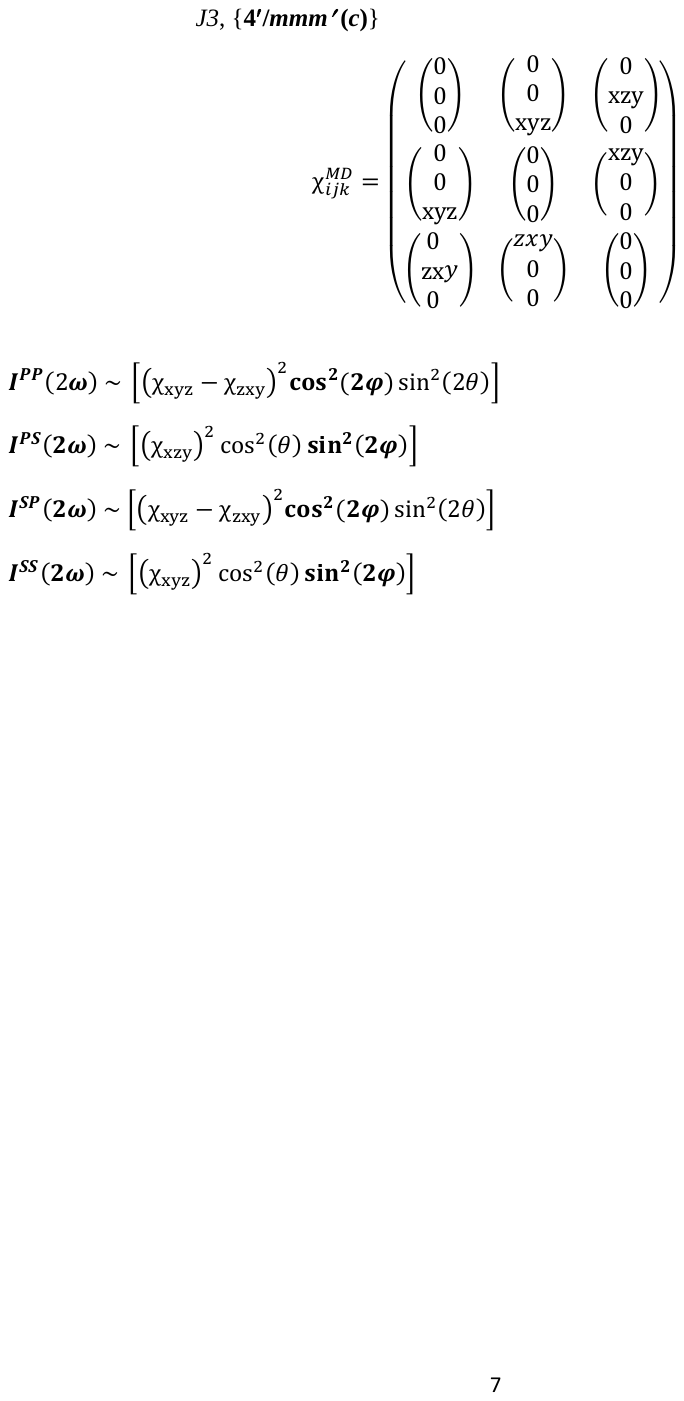}
    \caption{Caption}
    \label{fig:my_label}
\end{figure}
\begin{figure}
   \vspace*{-2cm}
   \hspace*{-2cm}
    \centering
    \includegraphics[page=8]{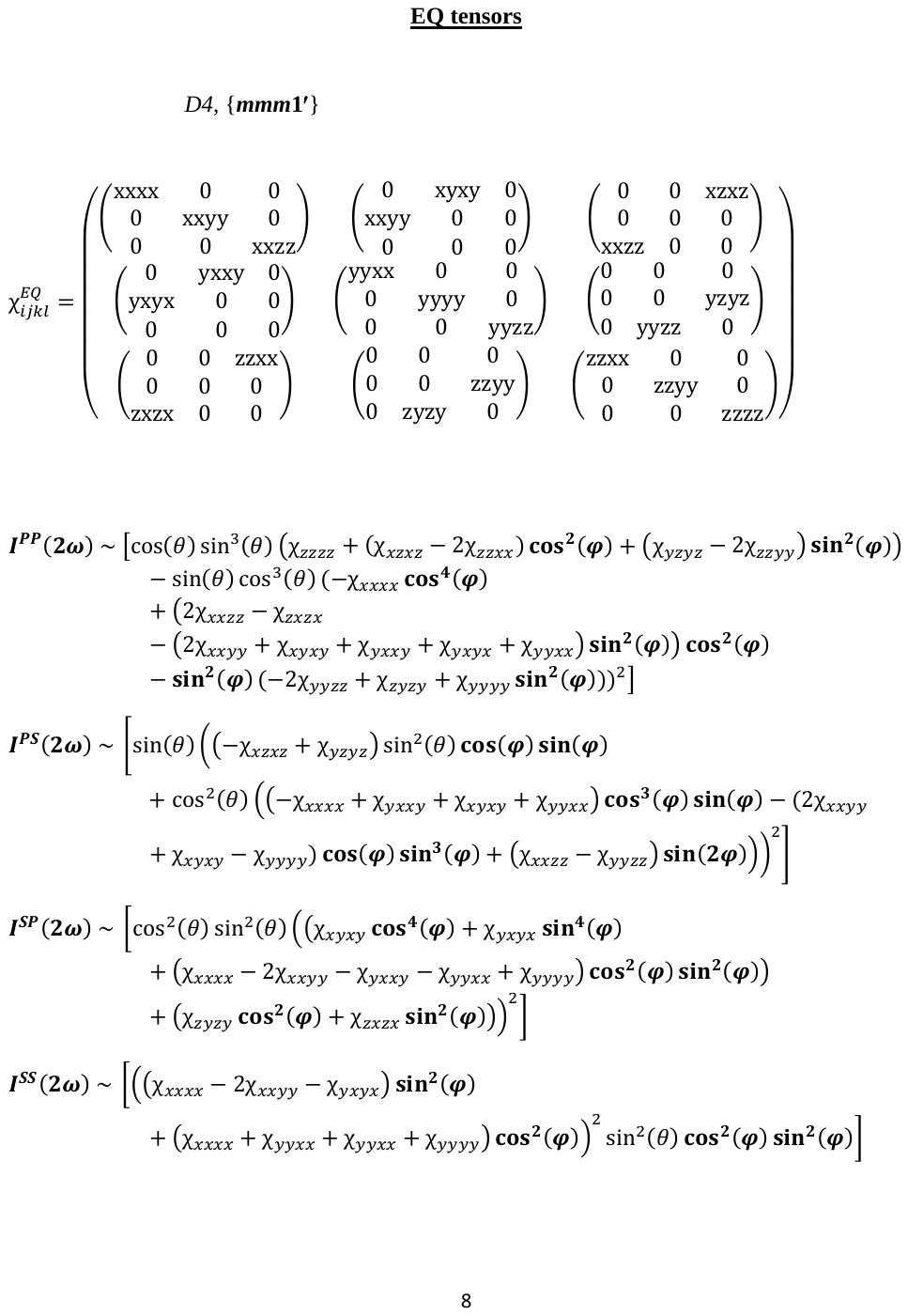}
    \caption{Caption}
    \label{fig:my_label}
\end{figure}\begin{figure}
   \vspace*{-2cm}
   \hspace*{-2cm}
    \centering
    \includegraphics[page=9]{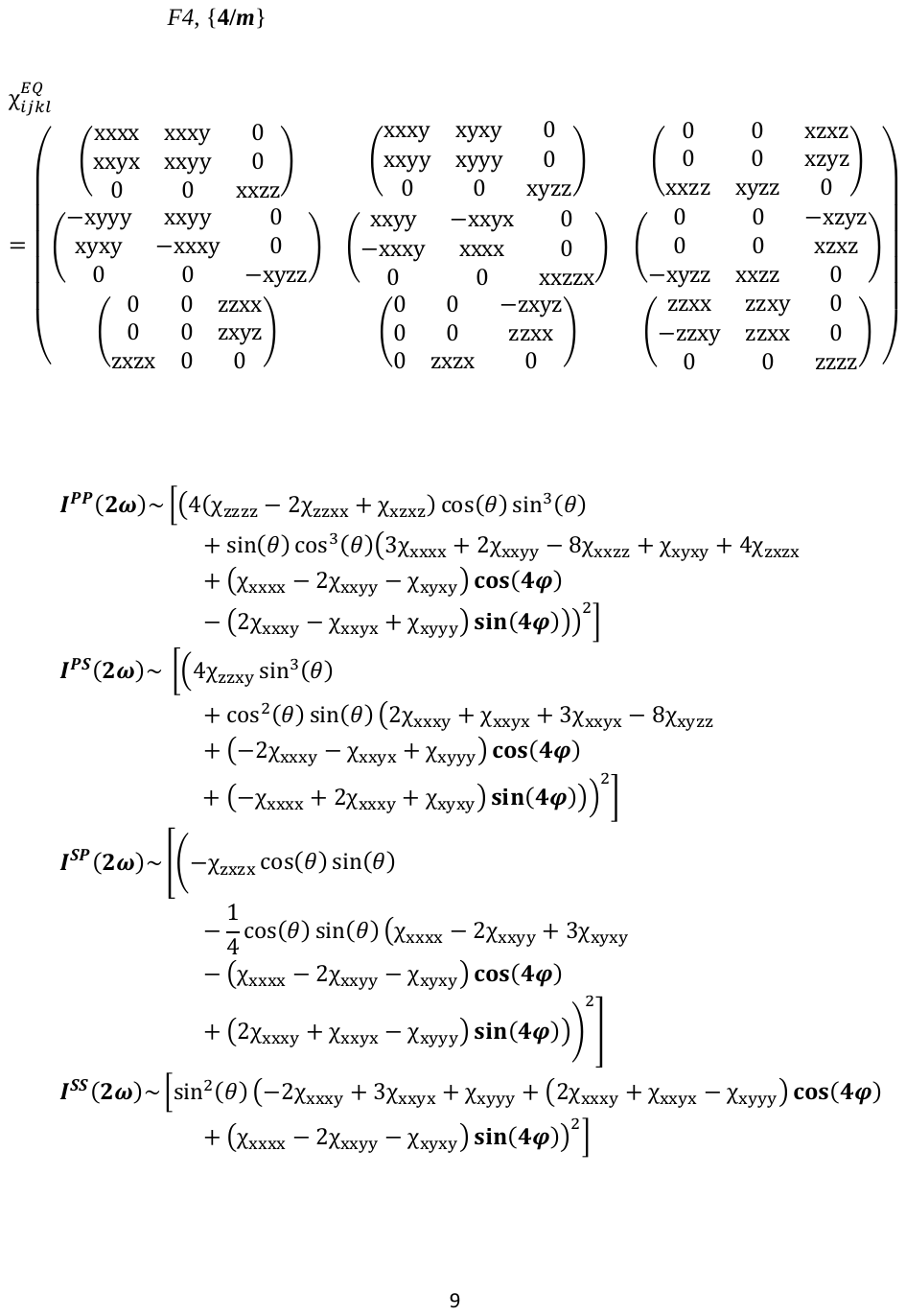}
    \caption{Caption}
    \label{fig:my_label}
\end{figure}
\begin{figure}
   \vspace*{-2cm}
   \hspace*{-2cm}
    \centering
    \includegraphics[page=10]{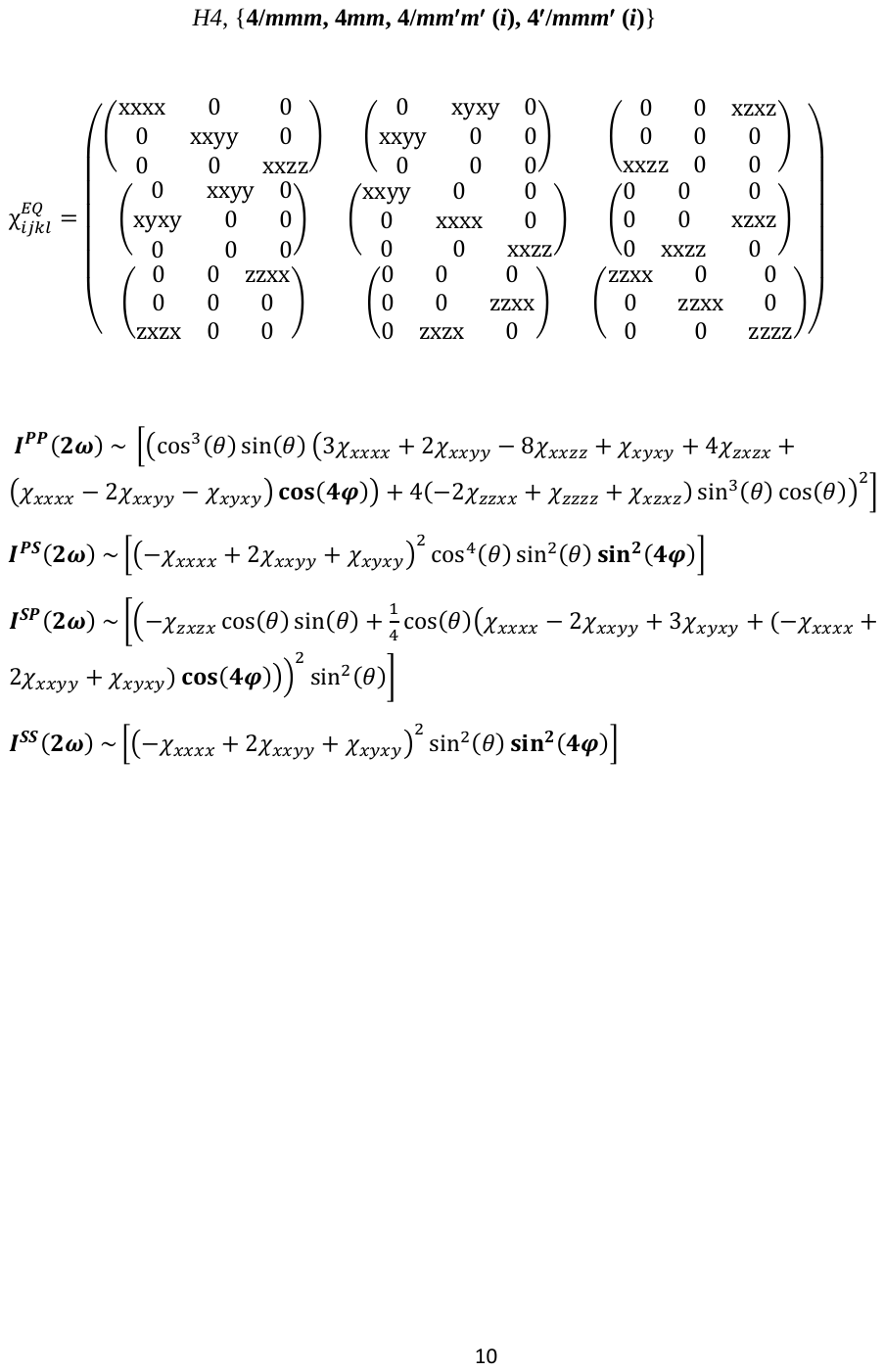}
    \caption{Caption}
    \label{fig:my_label}
\end{figure}
\begin{figure}
   \vspace*{-2cm}
   \hspace*{-2cm}
    \centering
    \includegraphics[page=11]{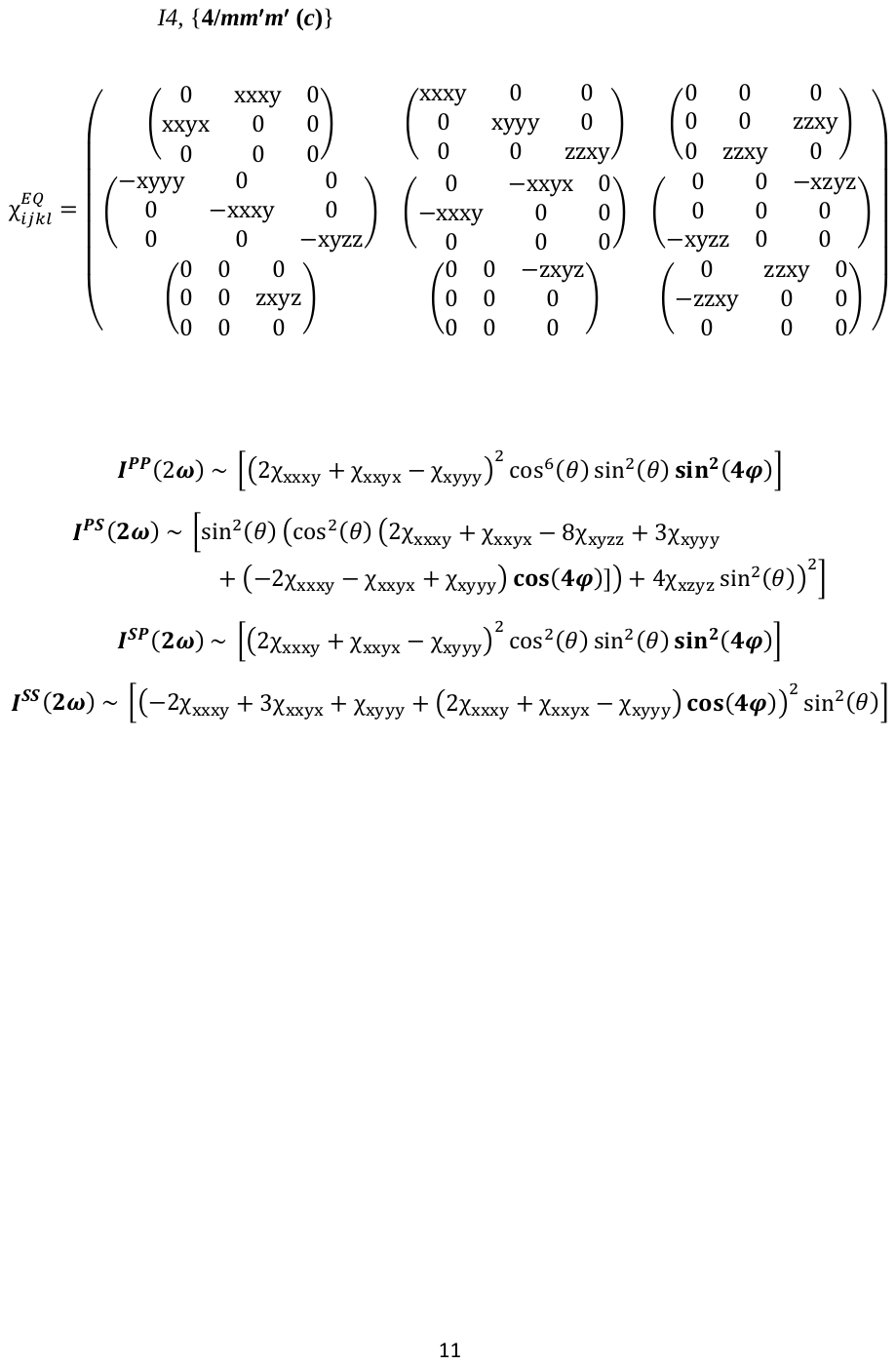}
    \caption{Caption}
    \label{fig:my_label}
\end{figure}
\begin{figure}
   \vspace*{-2cm}
   \hspace*{-2cm}
    \centering
    \includegraphics[page=12]{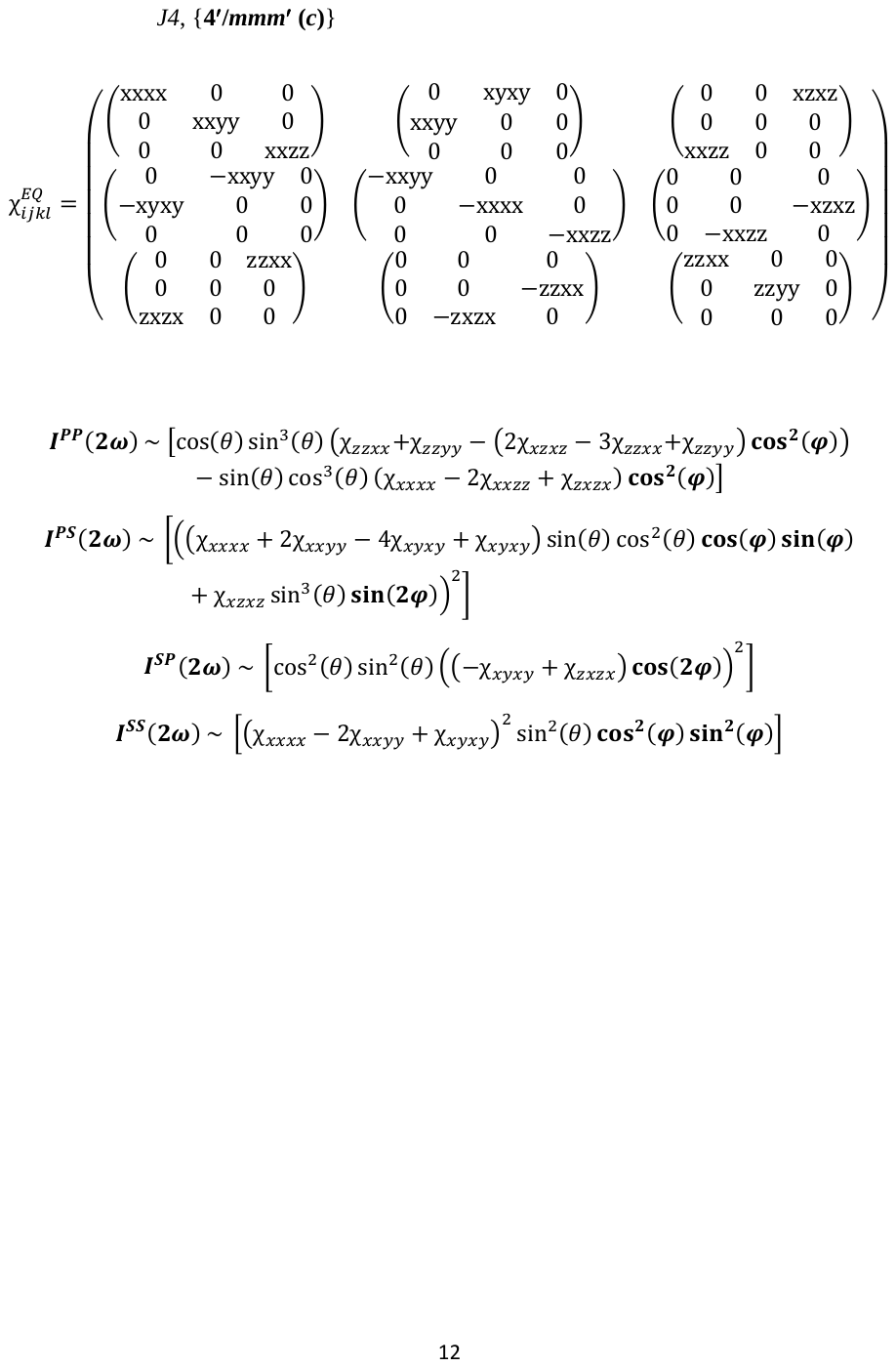}
    \caption{Caption}
    \label{fig:my_label}
\end{figure}
\begin{figure}
   \vspace*{-2cm}
   \hspace*{-2cm}
    \centering
    \includegraphics[page=13]{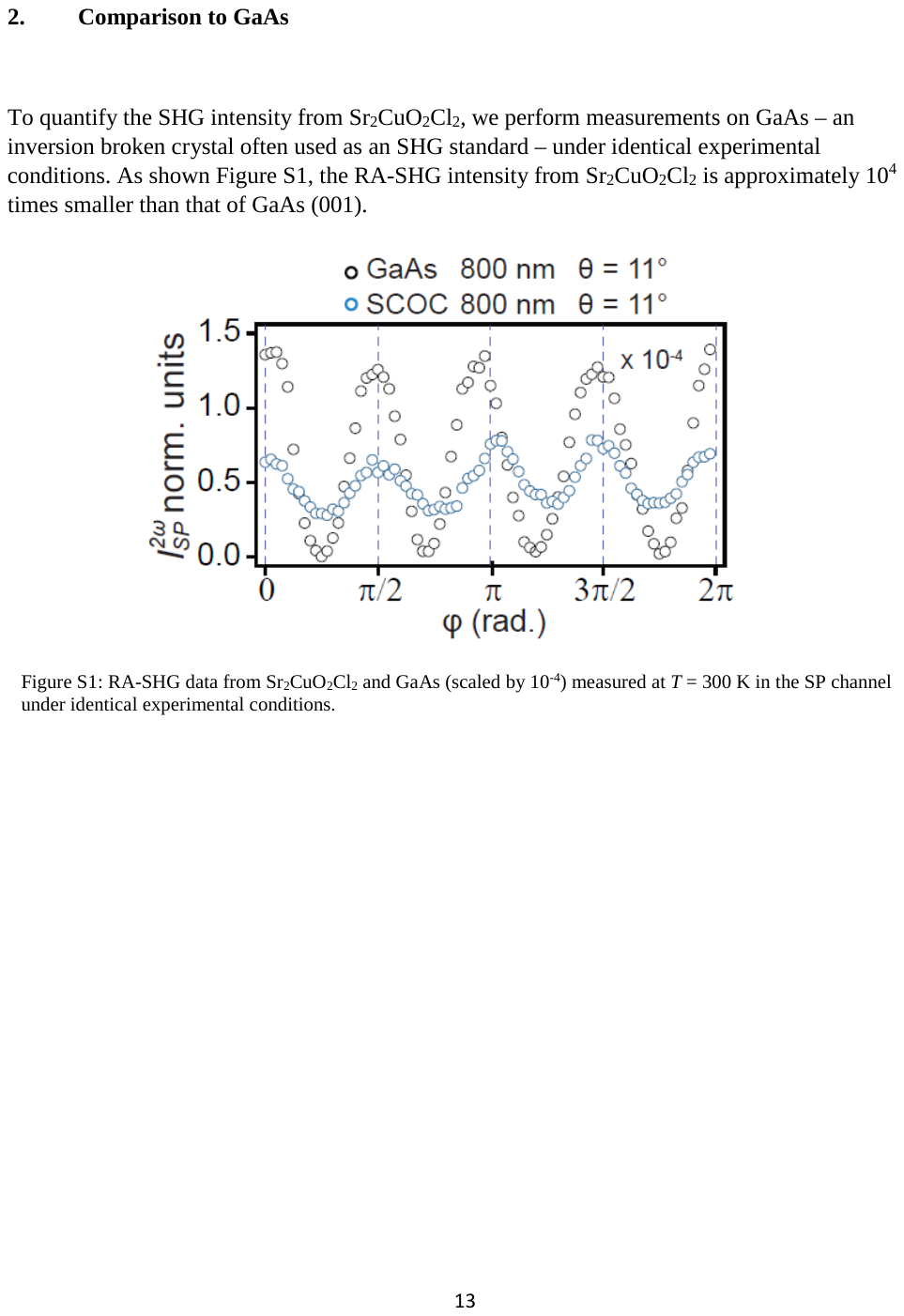}
    \caption{Caption}
    \label{fig:my_label}
\end{figure}
\begin{figure}
   \vspace*{-2cm}
   \hspace*{-2cm}
    \centering
    \includegraphics[page=14]{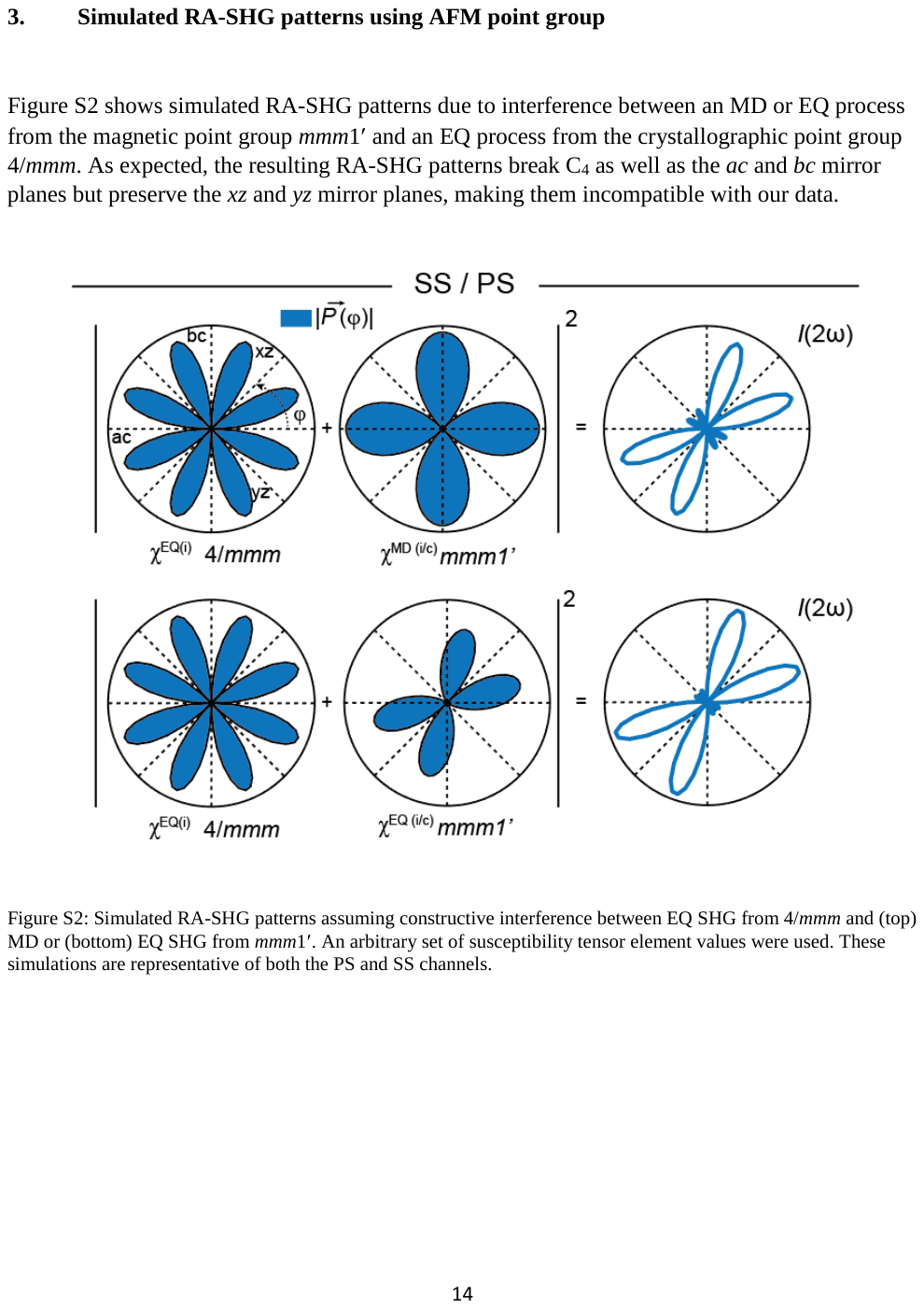}
    \caption{Caption}
    \label{fig:my_label}
\end{figure}
\begin{figure}
   \vspace*{-2cm}
   \hspace*{-2cm}
    \centering
    \includegraphics[page=15]{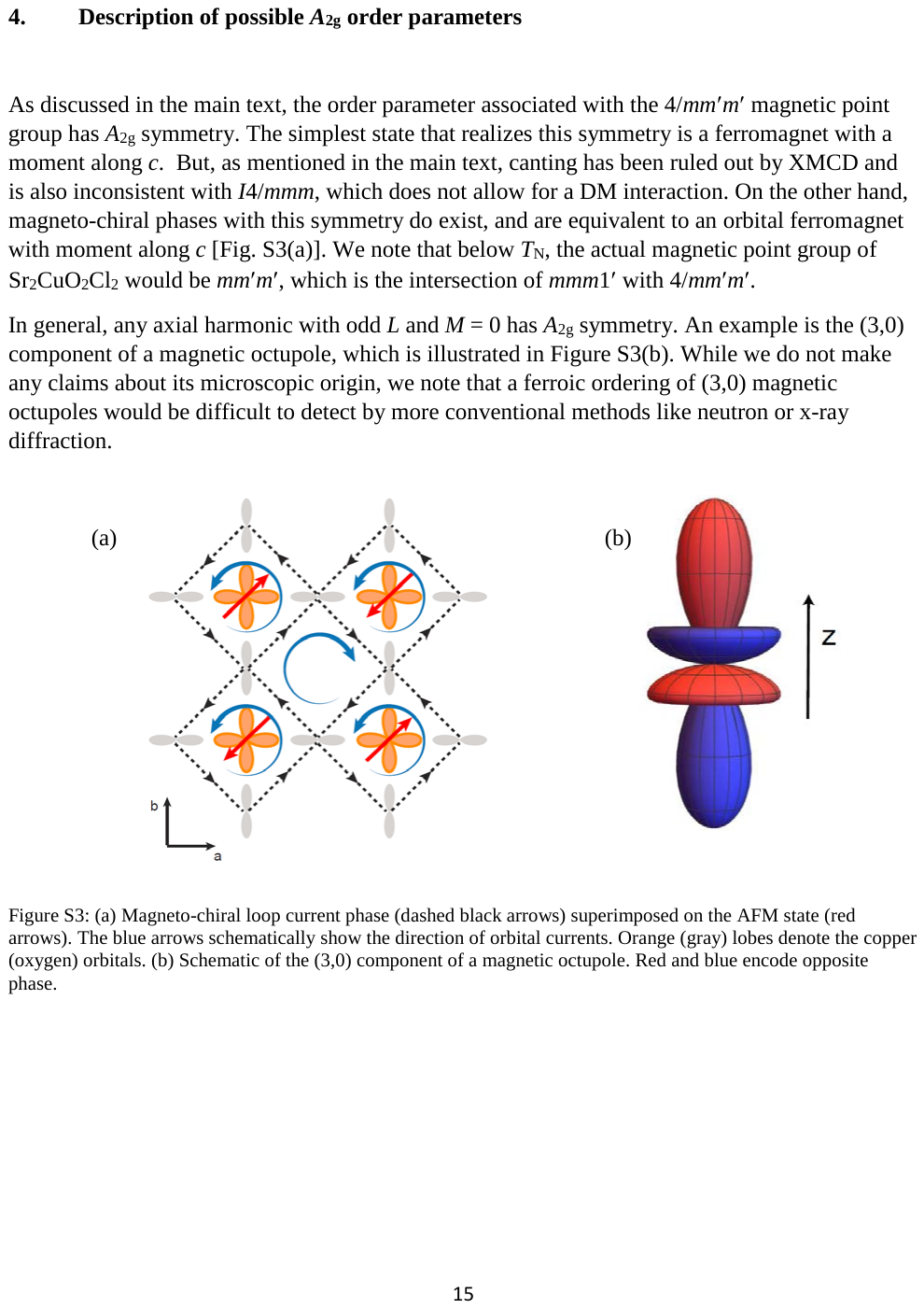}
    \caption{Caption}
    \label{fig:my_label}
\end{figure}
\begin{figure}
   \vspace*{-2cm}
   \hspace*{-2cm}
    \centering
    \includegraphics[page=16]{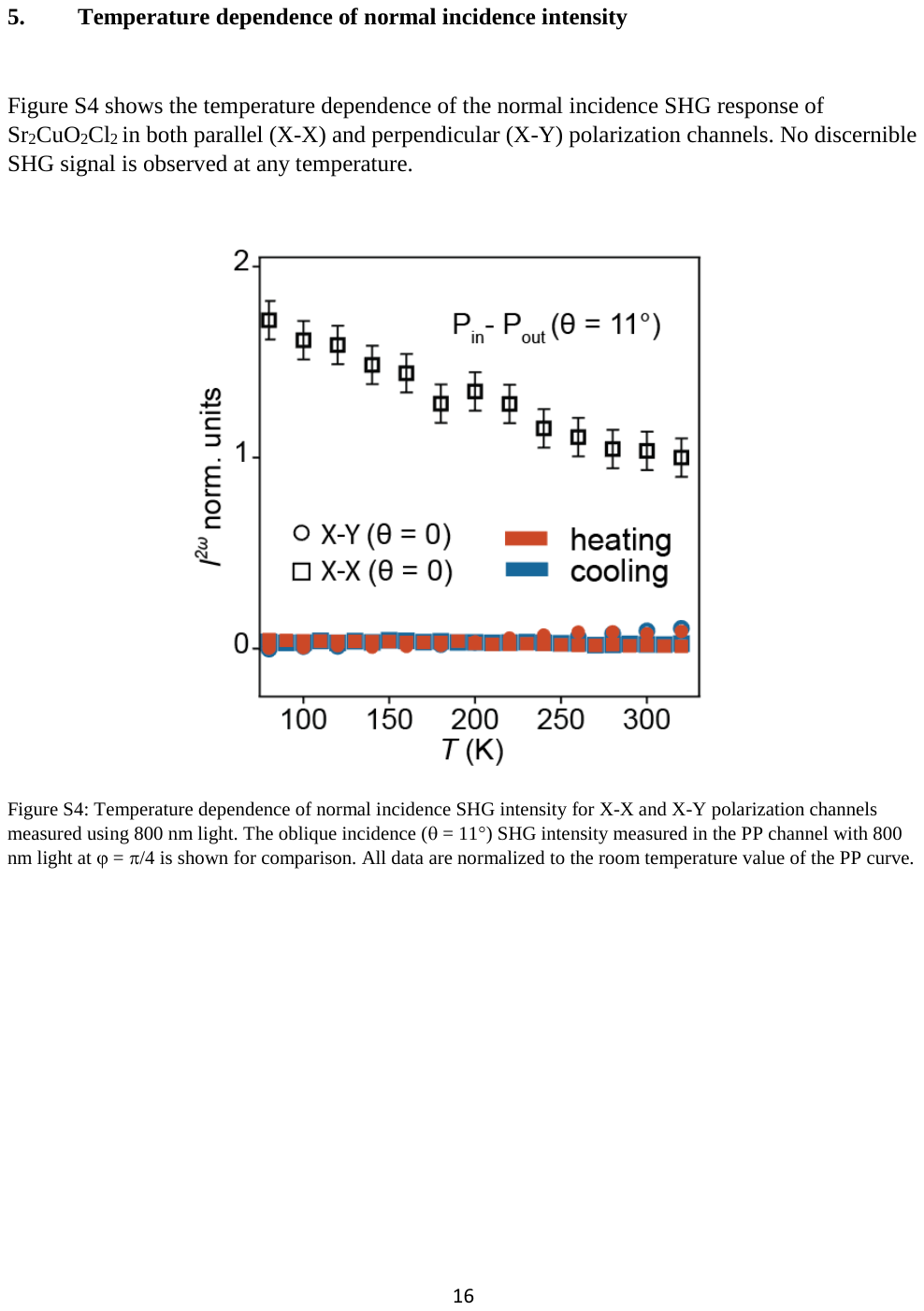}
    \caption{Caption}
    \label{fig:my_label}
\end{figure}
\begin{figure}
   \vspace*{-2cm}
   \hspace*{-2cm}
    \centering
    \includegraphics[page=17]{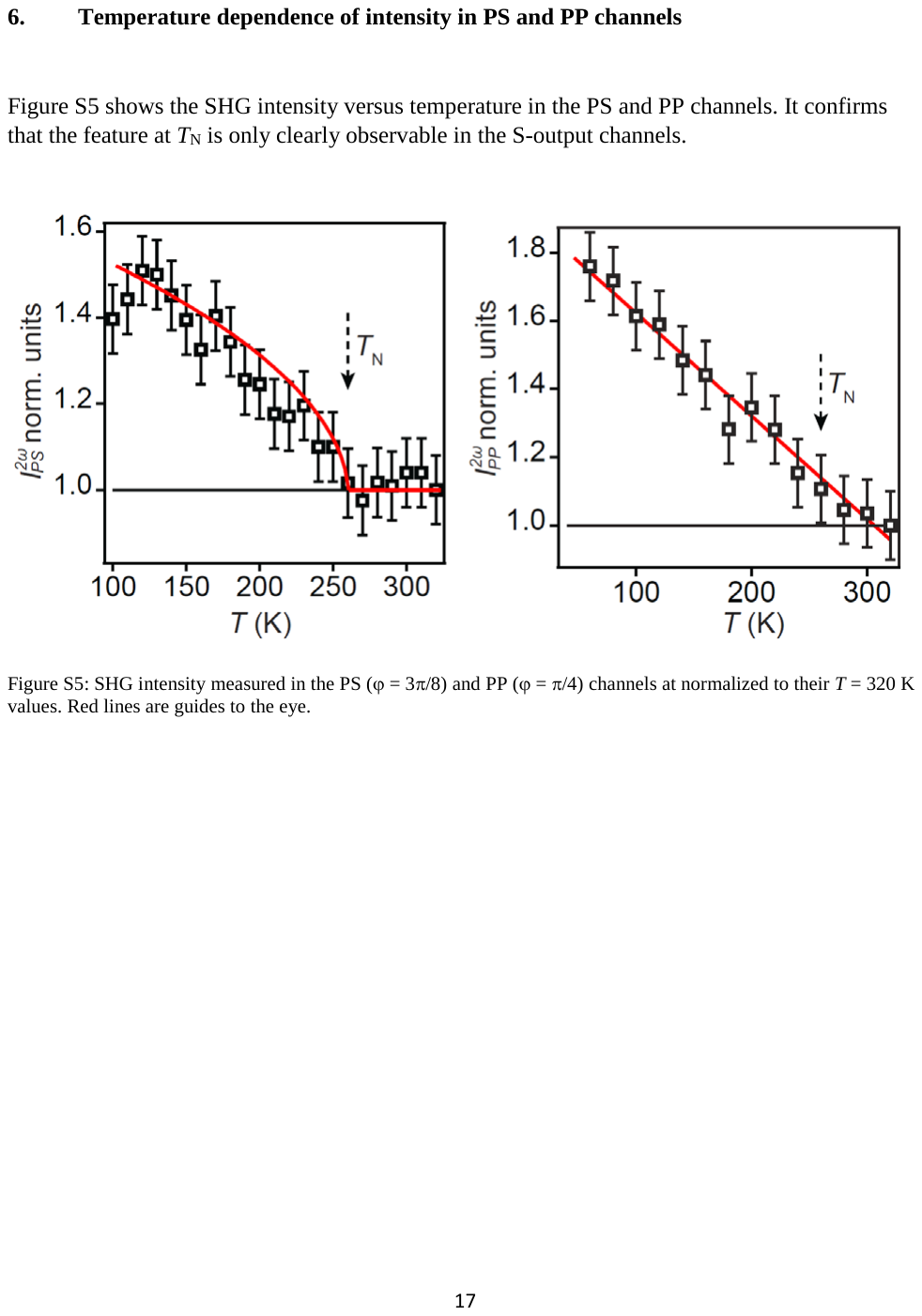}
    \caption{Caption}
    \label{fig:my_label}
\end{figure}
\begin{figure}
   \vspace*{-2cm}
   \hspace*{-2cm}
    \centering
    \includegraphics[page=18]{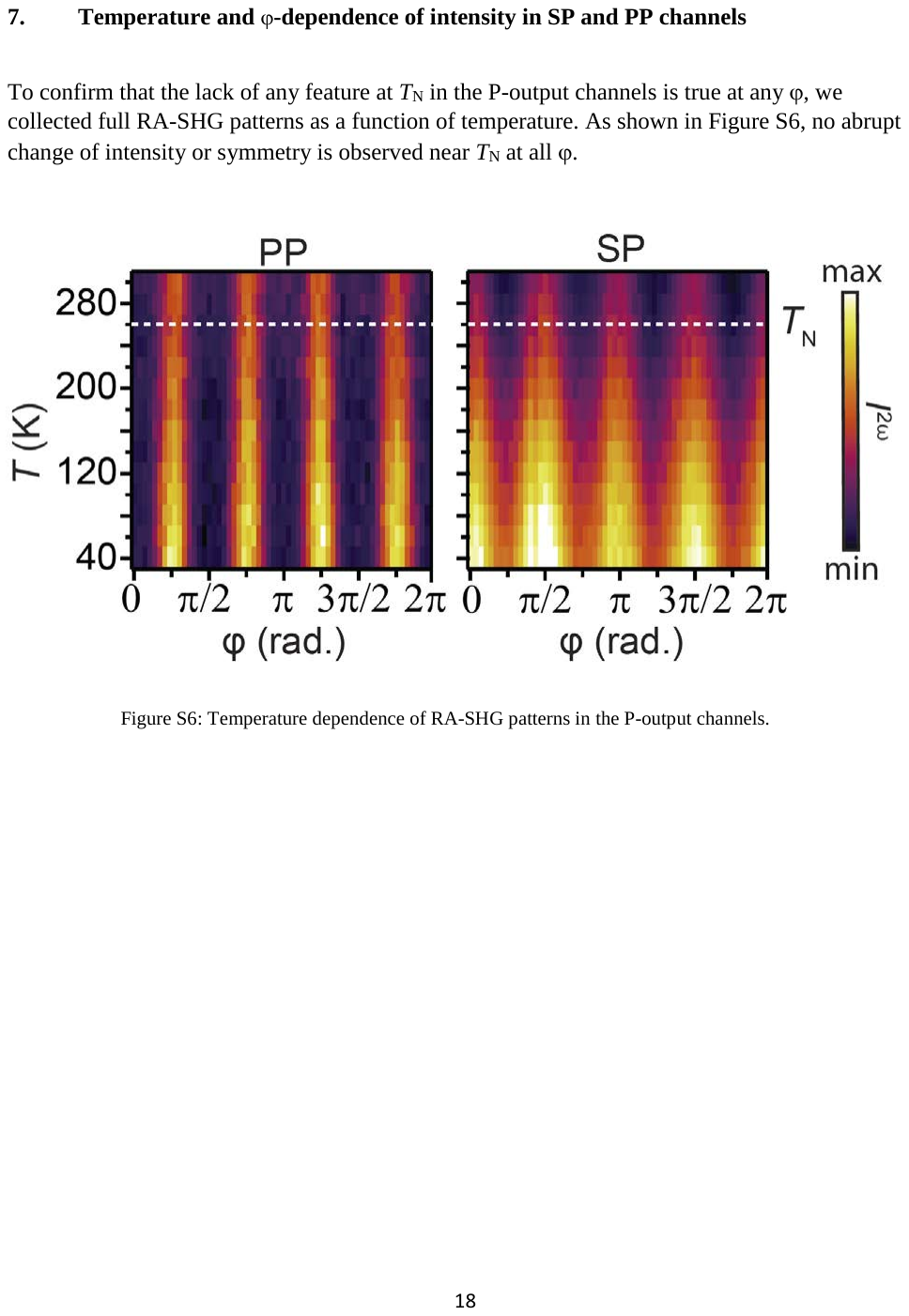}
    \caption{Caption}
    \label{fig:my_label}
\end{figure}
\begin{figure}
   \vspace*{-2cm}
   \hspace*{-2cm}
    \centering
    \includegraphics[page=19]{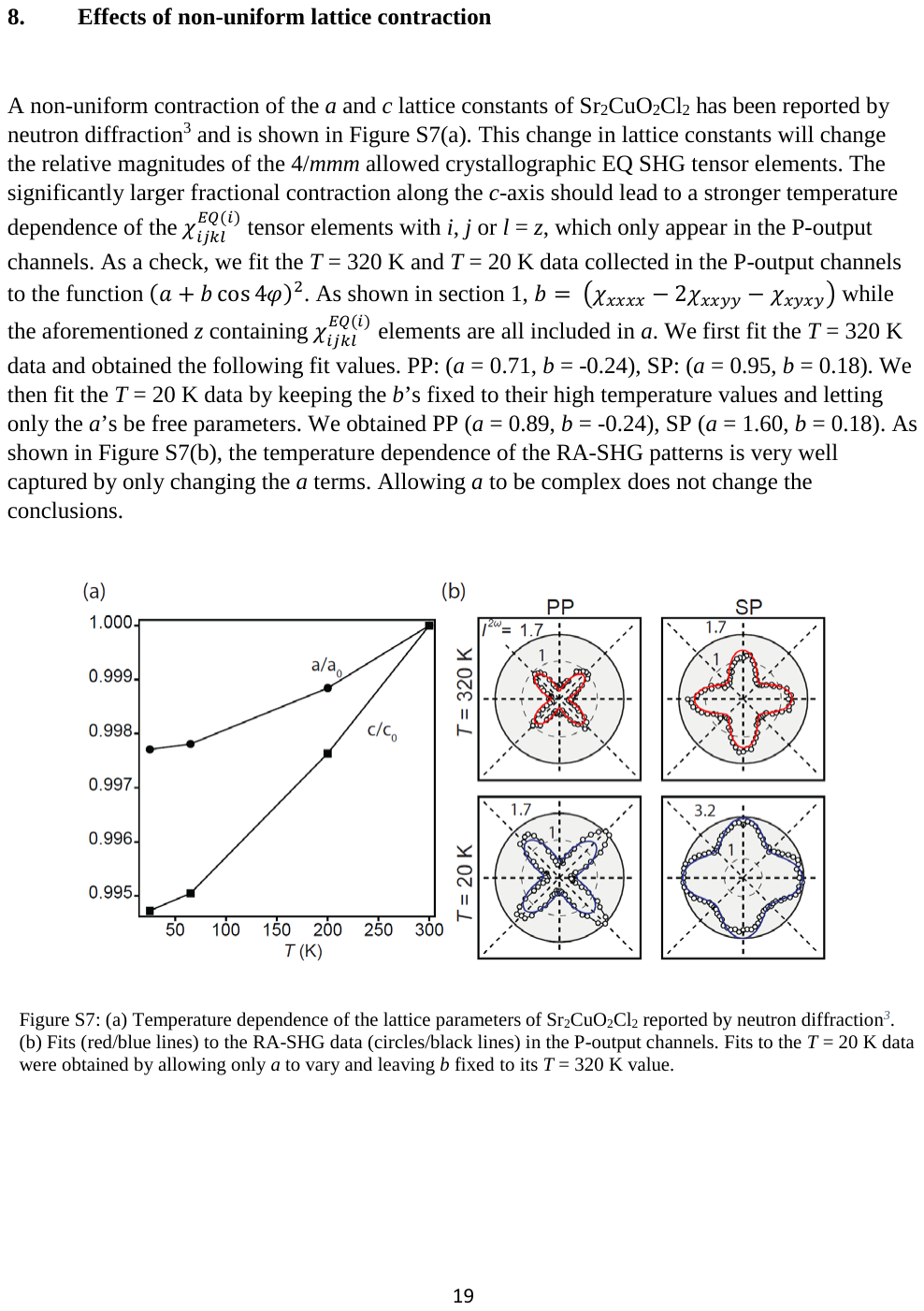}
    \caption{Caption}
    \label{fig:my_label}
\end{figure}
\begin{figure}
   \vspace*{-2cm}
   \hspace*{-2cm}
    \centering
    \includegraphics[page=20]{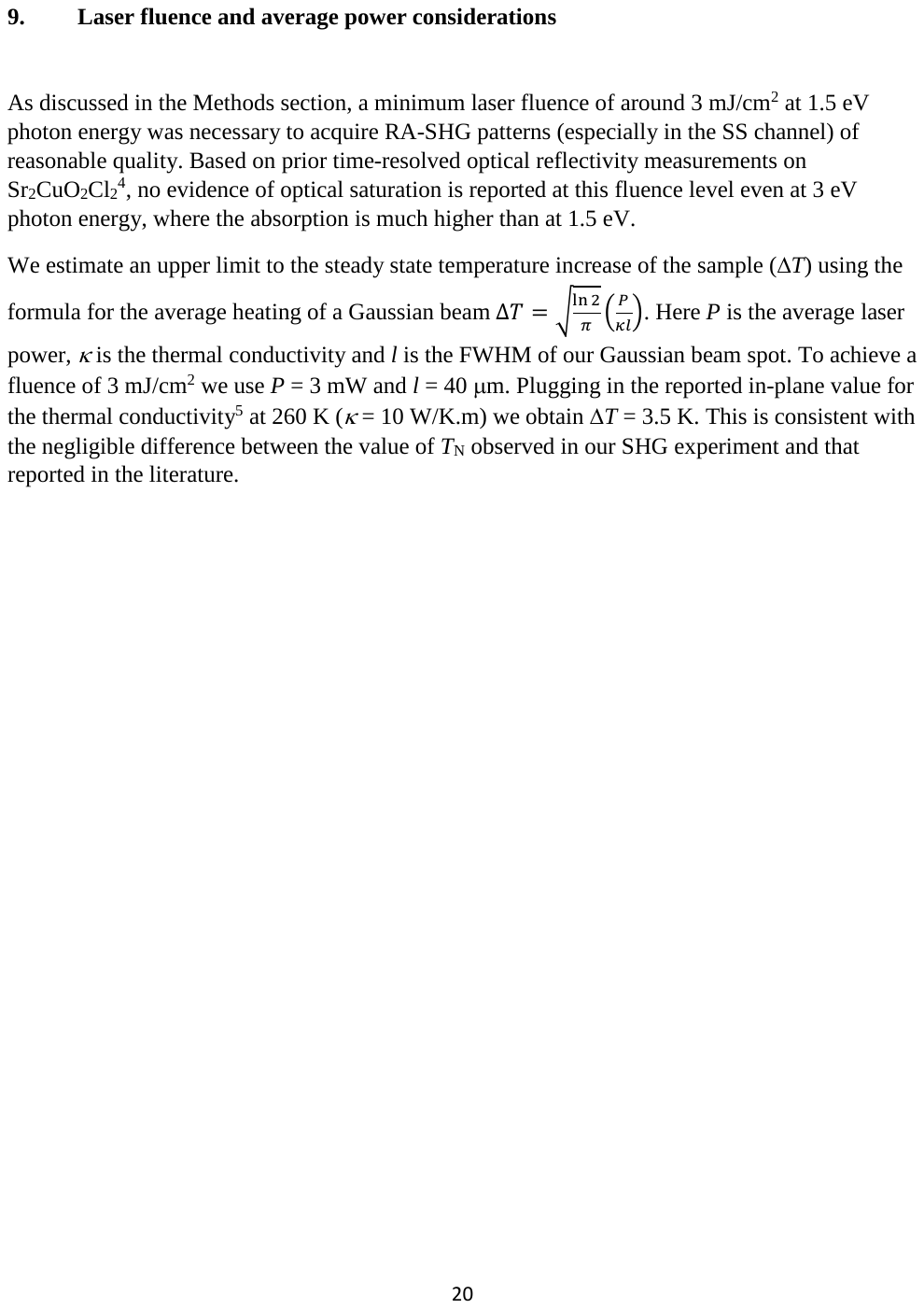}
    \caption{Caption}
    \label{fig:my_label}
\end{figure}
\begin{figure}
   \vspace*{-2cm}
   \hspace*{-2cm}
    \centering
    \includegraphics[page=21]{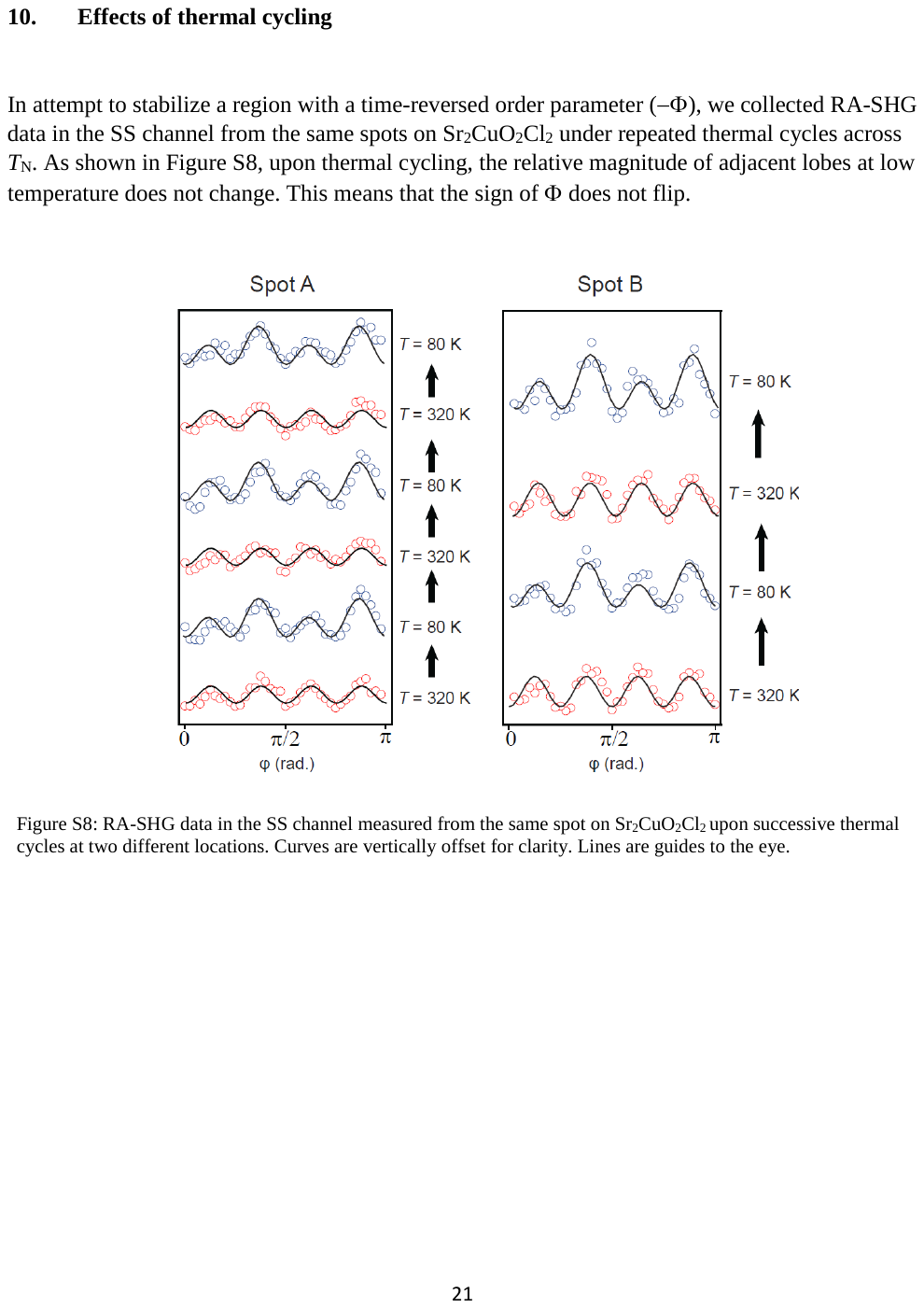}
    \caption{Caption}
    \label{fig:my_label}
\end{figure}
\begin{figure}
   \vspace*{-2cm}
   \hspace*{-2cm}
    \centering
    \includegraphics[page=22]{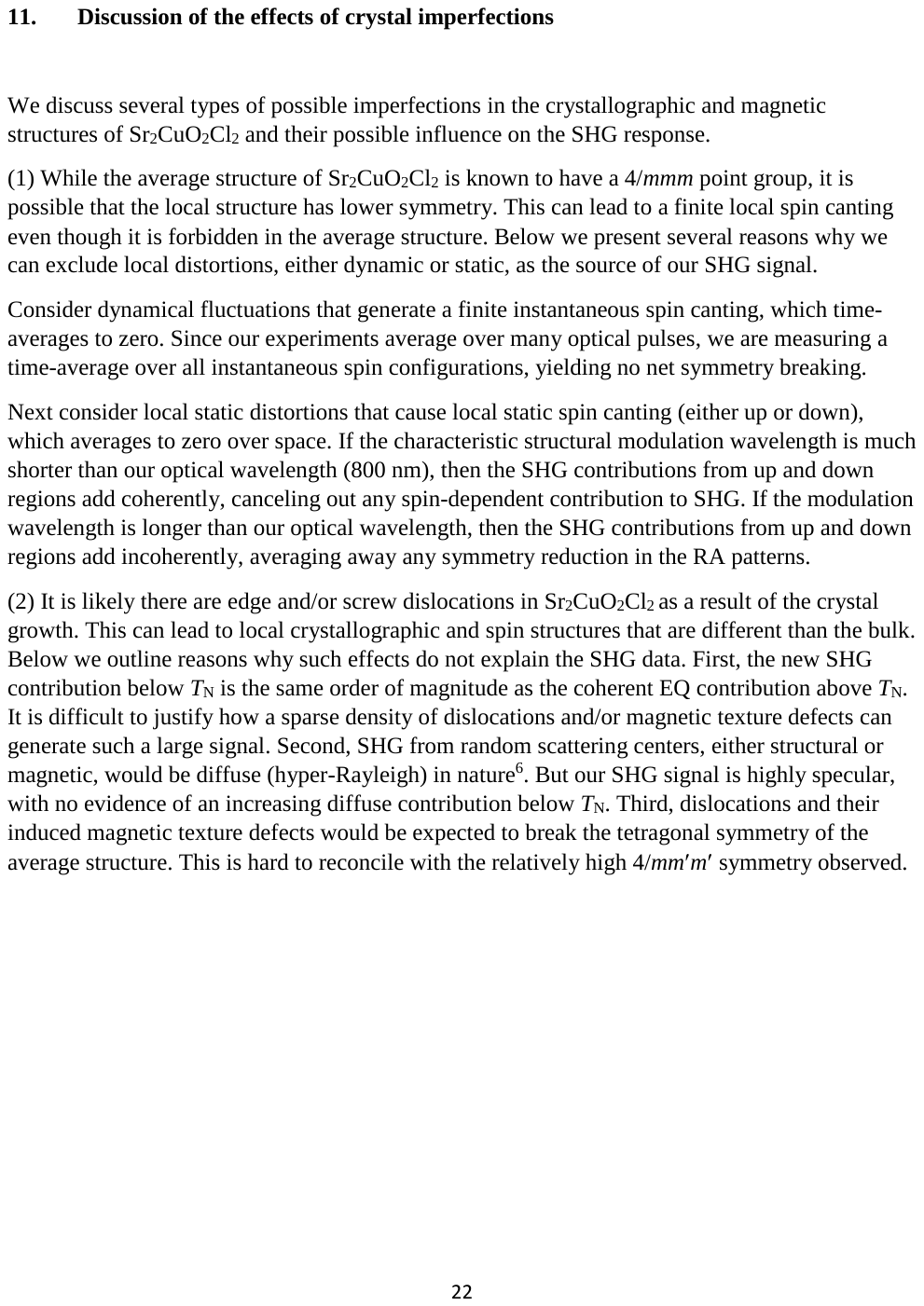}
    \caption{Caption}
    \label{fig:my_label}
\end{figure}
\begin{figure}
   \vspace*{-2cm}
   \hspace*{-2cm}
    \centering
    \includegraphics[page=23]{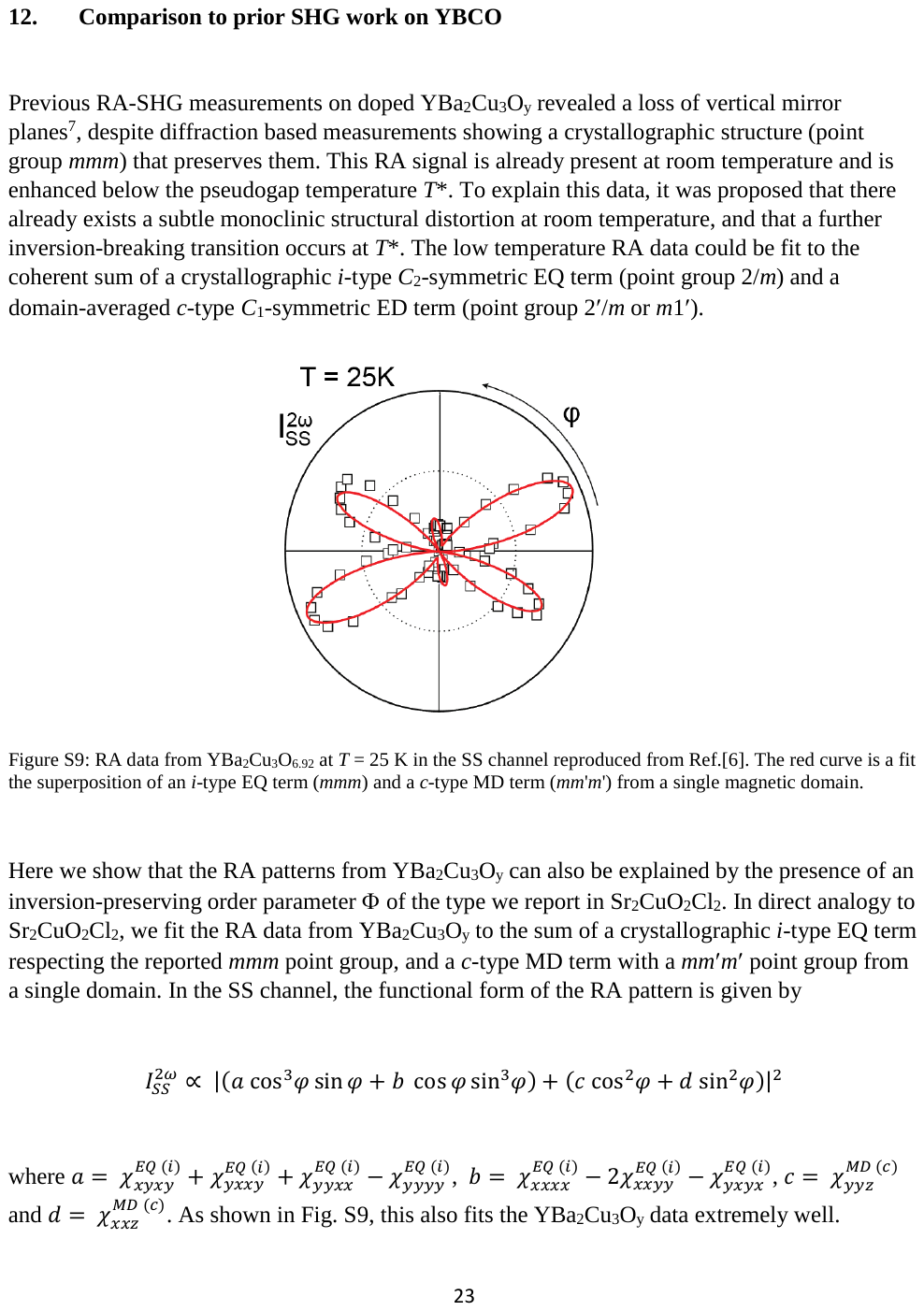}
    \caption{Caption}
    \label{fig:my_label}
\end{figure}
\begin{figure}
   \vspace*{-2cm}
   \hspace*{-2cm}
    \centering
    \includegraphics[page=24]{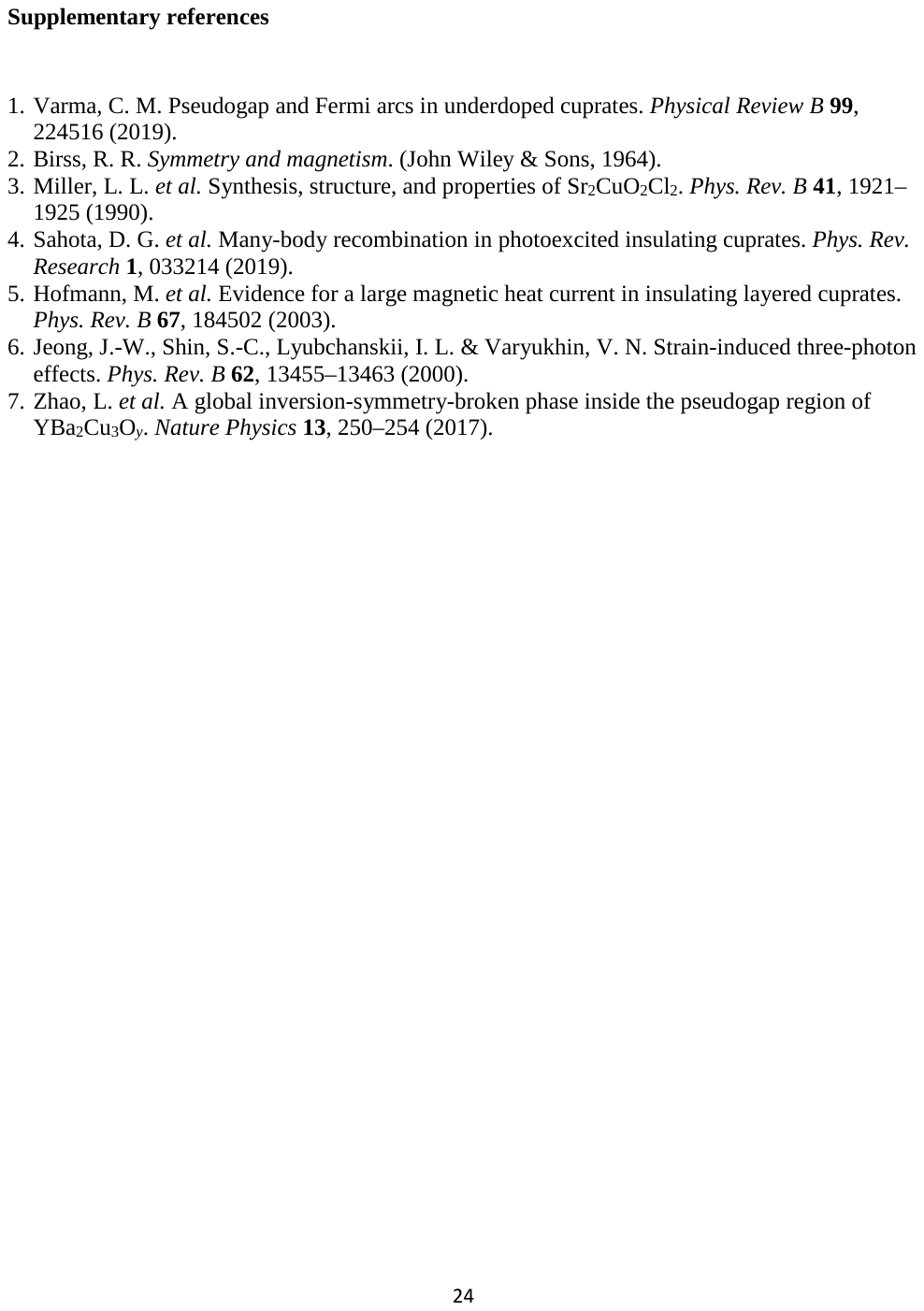}
    \caption{Caption}
    \label{fig:my_label}
\end{figure}
\end{document}